\documentclass{article}
\usepackage{multirow}

\usepackage{arxiv}

\usepackage[utf8]{inputenc} 
\usepackage[T1]{fontenc}    
\usepackage{hyperref}       
\usepackage{url}            
\usepackage{booktabs}       
\usepackage{amsfonts}       
\usepackage{nicefrac}       
\usepackage{microtype}      
\usepackage{lipsum}		
\usepackage{graphicx}
\usepackage{doi}
\usepackage{amsmath}
\usepackage{algorithm}
\usepackage[noend]{algpseudocode}
\usepackage{caption}
\usepackage{subcaption}
\usepackage{tikz}
\usepackage{enumitem}
\usepackage[normalem]{ulem}

\title{The Ubiquitous Skiplist: A Survey of What Cannot be Skipped About the Skiplist and its Applications in Big Data Systems}

\author{Lu Xing\\
	Department of Computer Science\\
	Purdue University\\
	West Lafayette, IN 47906 \\
	\texttt{vvadrevu@purdue.edu} \\
	\And
	Venkata Sai Pavan Kumar Vadrevu\\
	Department of Computer Science\\
	Purdue University\\
	West Lafayette, IN 47906 \\
	\texttt{xingl@purdue.edu} \\
	\AND
        Walid G. Aref\\
	Department of Computer Science\\
	Purdue University\\
	West Lafayette, IN 47906 \\
	\texttt{aref@purdue.edu} \\
}

\date{}

\renewcommand{\shorttitle}{What Cannot be Skipped About the Skiplist}

\begin{document}
\maketitle

\begin{abstract}
Skiplists have become  prevalent in systems. 
  The main advantages of skiplists are their simplicity and ease of implementation, and the ability to support operations in the same asymptotic complexities as their tree-based counterparts. 
  In this survey, we explore skiplists and their many variants. We highlight many scenarios about how skiplists are useful, and how they fit well in these usage scenarios. We also compare skiplists with other data structures, especially  tree-based structures. Extensions to skiplists include  structural modifications, as well as  algorithmic enhancements and  operations. We categorize the existing extensions, and summarize the skiplist variants that belong to each category. We present how data systems incorporate  skiplist variants into many different application scenarios to serve various purposes. These data systems cover a wide range of applications, from data indexing to block-chain, from network algorithms to deterministic skiplists, etc. 
  It illustrates an impactful and diverse applications of skiplists in various domains of data systems.
\end{abstract}

\keywords{Skiplist, skip list, index}

\section{Introduction}
\begin{figure}[h]
    \centering
    \begin{subfigure}{0.65\textwidth}
    \centering
        \includegraphics[width=\linewidth]{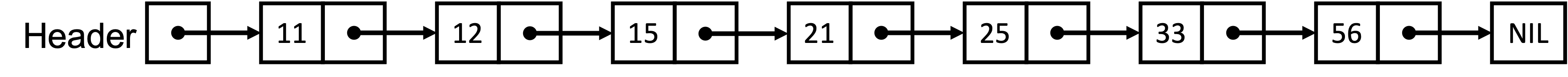}
        \caption{The linked list with seven keys.}\label{fig:linkedlist}
    \end{subfigure}
    \vfill
    \begin{subfigure}{0.65\textwidth}
    \centering
        \includegraphics[width=\linewidth]{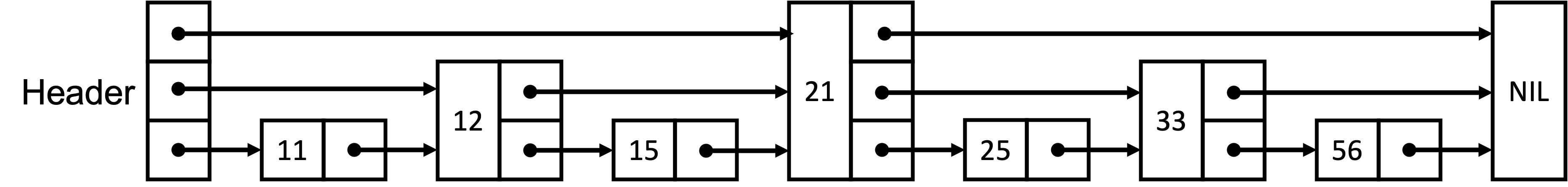}
        \caption{Each higher level  is a linked list that skips every other data item in the level below.}\label{fig:linkedlist_skip}
    \end{subfigure}
    \vspace{-3mm}
    \caption{The linked list and the skiplist that skips every other data item in the linked list below.}
\end{figure}

The linked list is a  popular and simple data structure~\cite{linkedlist}. 
It is composed of an ordered list of data items or nodes that are linked to one another. 
Each node has two fields: A {\em value} field that represents the value of the node, and a {\em next} pointer that points to the next node or data item in the linked list.
A {\em header} pointer points to the first node in the linked list.
Figure~\ref{fig:linkedlist} illustrates an example linked list. The last node in the linked list has a {\em NIL} pointer to designate the end-of-list. 

One drawback of the linked list is its search performance. 
Searching for a value  involves scanning the list from begin to end,  and hence is in the order of $O(n)$, where $n$ is the number of nodes in the linked list. An interesting  question arises: How can one speed up search in the linked list?

One  interesting solution to the search performance in the linked list has been introduced by William Pugh~\cite{pugh1990skip} in what he terms the {\em Skiplist}. What if we create a hierarchy of linked lists, where the bottom level of the hierarchy contains the full linked list containing all data items. Then, 
in each level up in the hierarchy, we create a new linked list that starts from the same head data item of the original linked list, but skip every other data item in the linked list in the level below and point to the item that follows in the linked list. Refer to Figure~\ref{fig:linkedlist_skip}. The data items that are repeated across levels of the hierarchy are interconnected so that one can navigate across the multiple linked lists in the hierarchy. The repeated interconnected data items across levels provide the ability to jump from one linked listed in the hierarchy to the next. 
The resulting data structure forms the basis of the {\em Skiplist}~\cite{pugh1990skip}. 
The top levels of the  skiplist have fewer data items, and hence can help speed up the search by {\em skipping} over the irrelevant items as we explain below.

There are many variants of the skiplist. The variant described above is  the {\em deterministic} skiplist. Searching for a data item in the skiplist takes $O(\log n)$ time. Other skiplist variants exist that address emerging challenges. We cover these variants and the challenges they address in this paper.
In contrast to the deterministic skiplist, the original skiplist, as introduced in~\cite{pugh1990skip}, is probabilistic.

In a traditional binary tree,  
when data is skewed, some sequences of data inserts may produce a degenerate binary tree that has poor performance. Although random permutations of the order of the data to be inserted can alleviate this issue, it is not feasible as queries are often demanded in real-time.
Being probabilistic, the original skiplist~\cite{pugh1990skip} 
addresses the issue of data skew, and is highly likely to produce a better structure that is better balanced regardless of the input data order.

The skiplist can be viewed as an alternative to balanced trees. The skiplist balances the structure probabilistically, which is easier than explicitly and intentionally maintaining the balance~\cite{pugh1990skip}. The simplicity of the skiplist makes it  attractive for 
adoption in systems. The expected search cost in the skiplist is the same as 
that of a 
balanced 
search 
tree. 
Moreover, the 
skiplist does not need re-balancing, and hence its   realization in systems and optimizing its performance are simpler.

Over the years, data indexes have been designed and deployed in systems, e.g., the B/B$^{+}$-tree~\cite{bayer1970organization,comer1979ubiquitous}, hash indexing~\cite{maurer1975hash}, the R-tree~\cite{guttman1984r}, the Log-Structured Merge Tree~\cite{o1996log}, and many of their variants. 
In recent years, the skiplist has become ubiquitous, not only as a probabilistic data structure that is of theoretical interest, but 
also
as a practical data index  that is being used widely in many systems.
The main advantages of skiplists are their simplicity and ease of implementation, and the ability to support operations in the same asymptotic complexities as their tree-based counterparts. 
Being an alternative to balanced trees, skiplists have become a candidate for database indexing. Furthermore, because of its distinct advantages and the extensibility of versatile optimizations, the skiplist 
is being used
in database 
systems, in 
distributed systems, and in networking protocols. 

In this paper, we survey the skiplist, its variants, and its applications in various data systems.
In Section~\ref{sec:basics}, we present basics of the skiplist, its structure, and its insert, delete, and search algorithms. Section~\ref{sec:contrast-btree} compares and contrasts the skiplist  with  other data structures, including the AVL tree, the self-adjusting tree, and the B/B$^+$-tree. 
Section \ref{sec:skiplist-opt} overviews potential optimizations that 
have been
applied to skiplists, and enumerates their corresponding skiplist variants.
Section~\ref{sec:taxonomy} introduces a  taxonomy for the various skiplists and their versatile uses  in various big data systems. 
Sections~\ref{sec:hardware}-\ref{sec:more-use} present applications of skiplist in various domains.
Finally, Section~\ref{section:conclude} concludes the paper.

\section{skiplist basics}\label{sec:basics}

In this section, we overview the basics of the skiplist. We review its structure, its insert, delete, and search operations along with analysis of the cost associated with each operation. Then, we present  common extensions and optimizations associated with the basic skiplist. In Section~\ref{sec:taxonomy}, we explain these extensions in more detail as we discuss variants of the skiplist.

\subsection{Basic Structure}\label{subsec:basic-structure}

\begin{figure*}[h]
\centering
    \begin{subfigure}{0.75\textwidth}
    \centering
        \includegraphics[width=\linewidth]{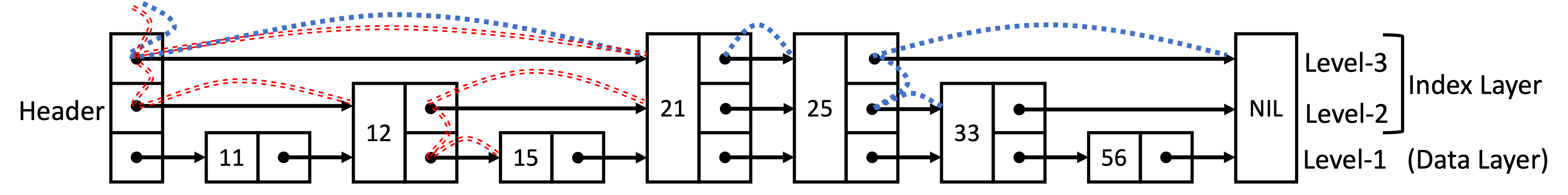}
        \caption{Searching for  15 and 33 in the skiplist}\label{fig:skiplist_search}
    \end{subfigure}
    \vfill
    \begin{subfigure}{0.75\textwidth}
    \centering
        \includegraphics[width=\linewidth]{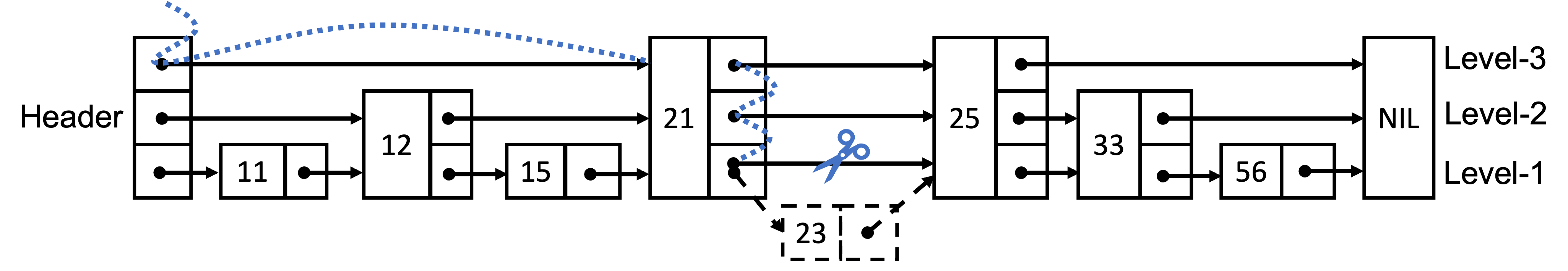}
        \caption{Inserting 23 into the skiplist}\label{fig:skiplist_insert}
    \end{subfigure}
    \vfill
    \begin{subfigure}{0.75\textwidth}
    \centering
        \includegraphics[width=\linewidth]{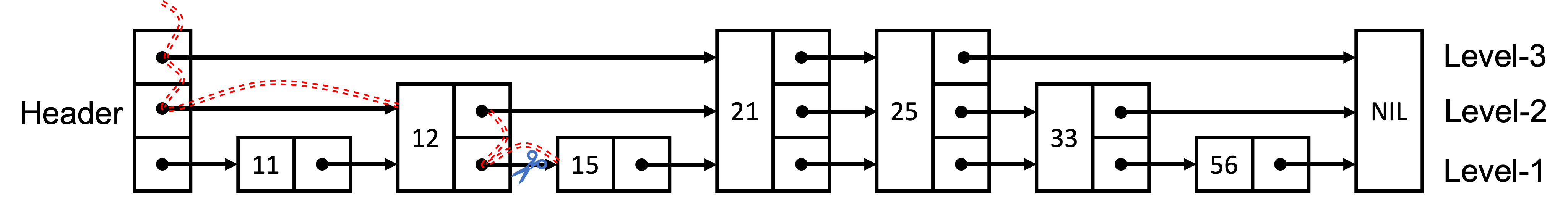}
        \caption{Deleting 15 from the skiplist}\label{fig:skiplist_delete}
    \end{subfigure}
\caption{The skiplist and its basic operations. (a) A skiplist with 7 keys, where the header and Nodes 21 and 25 have three forward pointers. The search path for Key 33 is illustrated by the blue dotted line; the search path for Key 15 is illustrated by the red double dashed line. (b)~To insert a new key 23, Node 21 modifies its forward pointers in Level-1. (c)~To delete 15, Node 12 modifies its forward pointers in Level-1. }
\end{figure*}

Refer to Figure~\ref{fig:skiplist_search} for illustration. The skiplist is a probabilistic data structure that consists of multiple levels, where each level contains a linked list of nodes and pointers. Levels are numbered bottom up from one to $i$ $(i \ge 1)$. Level-1 is a linked list connecting all the data items in sorted order. In 
Level-$i$ 
$(i>1)$, a fraction, say $p$, $(p\in (0,1))$ of the nodes in 
Level-$i$ 
also appear in 
Level-$(i+1)$.
Pugh~\cite{pugh1990skip}
uses $p=1/2$. Each node has at most one forward pointer per level pointing to its successor (right sibling) in the same level. The header of the skiplist has forward pointers; one per level. When the skiplist is empty, all forward pointers of the header point to NIL.  A node with $k$ forward pointers is referred to as a \textit{Level $k$} node. Given a node, its level is randomly chosen when it is inserted into the skiplist. The number of levels is capped by a constant \textit{MaxLevel} that Pugh~\cite{pugh1990skip} suggests to use $L(n) = \log_{1/p} N$ ($N$ is an upper bound of the number of elements in the skiplist). Figure~\ref{fig:skiplist_search} gives an example 
skiplist containing seven elements with $p=1/2$. Nodes 12 and 33  have two forward pointers. In contrast, the header and Nodes 21 and 25  have three forward pointers. We show how operations are conducted over the skiplist in the next section.

As in Figure~\ref{fig:skiplist_search}, a skiplist is logically divided into two parts: 
the index layers, and the data layer. Each query passes through the index layers, then reaches the data layer. The 
index layers 
route the query to the data layer, e.g., in $O(\log n)$ time. 
On the other hand, the data layer stores the actual data (key and  data), whereas the index layers only store key values used for routing purposes.

\subsection{Operations and Cost Analysis}\label{subsec:operations}

\subsubsection{The Search Operation}\label{subsubsec:search}
Search starts from the topmost level (maximum level available) and follows the forward pointers to reach the searched item. At each node, say $x$, the search key is compared with the node's key, say $x.y$. If $x.y$ is the same as the search key, then $x$'s value is returned. If $x.y$ is greater than the search key, the search process goes back to the previous node, moves down to the linked-list one level below, and repeats the same search process. In Figure~\ref{fig:skiplist_search}, if the search key is 33 (Follow the dotted blue line in Figure~\ref{fig:skiplist_search}), search  starts from the header, then to Nodes 21, 25, the search encounters \texttt{NIL} in the highest level as indicated  by the dotted blue line in Figure~\ref{fig:skiplist_search}, and thus the search goes one level down of Node 25, 
and finally reaches Node 33. If the search key is 15 (Follow the double dashed red line in Figure~\ref{fig:skiplist_search}), the search checks the highest level first. Since  Search Key 15 is smaller than 21, search goes one level down and checks Node 12, then Node 21, and then goes down to the lower level from Node 12, and finally reaches Node 15. 

The cost of search is proportional to the search path of the key. The expected average length of a search path is $O(\log n)$, where the list has $n$ elements. Detailed analysis is provided in Section~\ref{subsec:analysis-p}.

\subsubsection{The Insert Operation}\label{subsubsec:insert}

\begin{algorithm}
\caption{Determine Height of Node}\label{alg:random-level}
\begin{algorithmic}
    \Procedure{GetNodeHeight}{}
      \State $level \gets 1$
      \State Comment: $random()$ returns a random value in [0...1)
      \While{$random() < p$ and $level < MaxLevel$}
        \State $level = level + 1$
      \EndWhile
      \State {\bf return} $level$
    \EndProcedure
    \end{algorithmic}
\end{algorithm}

To insert a new data item, the height $h$ of the new node is determined by Algorithm~\ref{alg:random-level}. In Algorithm~\ref{alg:random-level}, the level is initially set to 1. The level is incremented (node is promoted by one level at a time) by consulting a random-number generator.
Then, the insert process starts by searching for the key in the skiplist following the search procedure in the previous section shown in dotted blue line in Figure~\ref{fig:skiplist_insert}. However, if the height of the key is greater than one, pointers need to be modified in multiple levels. The pointers of the rightmost node that is to the left of the inserted node in levels-$i$ ($i <= h$) are recorded during the search phase, and these pointers are updated to 
accommodate 
the insert. In Figure~\ref{fig:skiplist_insert}, suppose Key 23 is inserted and its height as returned by Algorithm~\ref{alg:random-level} is 1. After creating Node 23, the pointer of its level-1 predecessor is updated as in Figure~\ref{fig:skiplist_insert} bottom level. Thus, Node 21 points to Node 23 with the rest of the skiplist unchanged. The expected average cost is the same as that of search as it takes extra constant time to update pointers after the search.

\subsubsection{The Delete Operation}\label{subsubsec:delete}
Delete is similar to insert. When the search key is found (the double dashed red line in Figure~\ref{fig:skiplist_delete}), pointers are modified to reflect the delete, as  in Figure~\ref{fig:skiplist_delete} bottom level. If the delete key is 15, the forward pointer of its predecessor, i.e., Node 12, is modified in the bottom level, i.e., Level-1, to point to the successor of Node 15, which is Node 21.

\subsubsection{The Range Search Operation}\label{subsubsec:range-scan}
Given a range [\textit{start-key, end-key}], \textit{start-key} is searched for in the skiplist, then all the keys within the range can be found by following the bottom linked list.

\subsection{Analysis of {\em p}}\label{subsec:analysis-p}
Pugh~\cite{pugh1990skip} provides a detailed analysis of {\em p}. We summarize it below.
First, we analyze the expected cost of the search operation. 
By analyzing backwards of the search path,
we start at the bottom level 
of
the search key and traverse up and to the left.
Let $C(k)$ be the expected cost, i.e., the path length, that climbs up $k$ levels. There are two possibilities when the traversal determines where to go next at Level $k$, either left or upward. For example, if we are at Level-2's Node 33 in Figure~\ref{fig:skiplist_search}, 
the traversal continues to the left. But, if we are at Level-2's Node 21, 
we climb upwards, i.e., 
the traversal continues
upwards. The probability of climbing upwards is $p$ based on the definition, and is $1-p$ for traversing to the left. Thus, we have:
    $C(k) = (1-p)\times (C(k) + 1) + p\times C(k-1).$
Since $C(0)=0$, the simplified equation
is $C(k) = k/p$. The total expected cost to climb 
up
a skiplist of $n$ items is at most $L(n)/p + 1/(1-p)$. The constant value is due to the cost 
of
searching beyond $L(n)$.
The detailed analysis 
appears
in~\cite{pugh1990skip}.

Next, we calculate the average number of pointers per node 
and express that in terms of $p$.
Level-1 is a linked list of all $n$ items with $n$ pointers; Level-2 contains $p\times n$ items with $p\times n$ pointers; Level-3 contains $p^2\times n$ items with $p^2\times n$ pointers, etc. Thus, the average number of pointers per node is $1/(1-p)$, suggesting 
that, on average,
a larger $p$ requires more pointers per node.

Finally, we analyze the 
search time in  terms of $p$. Let $B(p)$ 
be
the expected search cost of a skiplist containing $n$ items. When $p = 1/2$, $B(1/2) = 2\log_2n$; when $p=1/4$, $B(1/4) = 4\log_4n=B(1/2)$; when $p=1/8$, $B(1/8) = 8\log_8n=4/3\times B(1/2)$; $p=1/16$, $B(1/16) = 16\log_{16}n=2\times B(1/2)$. The normalized search time increases with 
the
decrease in $p$, and so is the variability of running times. Pugh~\cite{pugh1990skip} 
states
that $p = 1/4$ slightly improves the constant factors. He suggests to use $p=1/4$ unless the variability in running times is a primary concern. Then, $p$ would be 1/2 in this case.

\section{Comparison with Other Data  Structures}\label{sec:contrast-btree}


In addition to the skiplist, other data structures can be used for the same purpose, and can perform in the same order bound. In this section, we compare the skiplist with these alternative data structures. 
The skiplist can be interpreted as a specific kind of tree~\cite{bose2008dynamic}. 
We focus 
primarily on the B/B$^{+}$-tree~\cite{bayer1970organization}. The B/B$^{+}$-tree is a popular  index structure. Thus, this comparison validates the skiplist as a potential candidte for serving as in  index in big data systems.

\subsection{
Comparing the Basic Skiplist Against the AVL  and the Self-Adjusting Trees
}\label{subsec:skiplist-vs-tree}

In terms of ease of implementation, the skiplist is easier to implement than balanced  and self-adjusting tree structures as the latter structures require special rebalancing code that is relatively complex. The skiplist does not require any rebalancing, and hence is much easier to implement~\cite{pugh1990skip}.

\subsubsection{
Performance Analysis}
This section compares the  skiplist against balanced trees~\cite{knuth1997art,wirth1976algorithms} and self-adjusting trees~\cite{sleator1985self}. 
\cite{pugh1990skip} 
compares 
implementations of 
various 
data structures while storing 2$^{16}$ integers
based on the relative timing of performing operations: 
The search time of the skiplist is set as the baseline. The non-recursive AVL tree takes 0.91$\times$, the recursive 2-3 tree takes 1.05$\times$, and the self-adjusting trees with top-down splaying takes 3.0$\times$, with bottom-up splaying taking 9.6$\times$. 
For insert time, the relative timings for the above data structures are 1.55$\times$, 3.2$\times$, 2.5$\times$, 7.8$\times$, respectively, relative to the skiplist. For  delete, the relative timings are similar to those for insert,  1.46$\times$, 3.65$\times$, 3.1$\times$, 9.0$\times$, respectively, relative to the skiplist.

Next, we compare the performance bounds of the above data structures~\cite{pugh1990skip}. The skiplist is  probabilistic   and has a probabilistic bound. In contrast, balanced trees have worst-case time bounds, and self-adjusting trees have amortized time bounds. If an operation needs to execute within a fixed time bound, self-adjusting trees may by undesirable~\cite{pugh1990skip}. For highly skewed data, self-adjusting trees are faster than skiplists~\cite{pugh1990skip} as  trees would adjust to the data distribution.

\subsubsection{Duality of the Skiplist}
A detailed transformation between the skiplist and the binary search tree (BST, for short) is discussed in~\cite{dean2007exploring}. A skiplist can be transformed into a multiway branching search tree with edge weights (weight indicates the node height difference between parent and child nodes), and then to a BST with edge weights. To achieve a balanced BST, tree rotation is involved. The promotion (demotion) of a node resulting from an insert or a delete into the skiplist is also performed by a rotation in the BST.

\subsection{Comparing the Deterministic Skiplist Against the B/B$^{+}-$ Tree}\label{subsec:deterministic-vs-tree}
The B-tree has been used ubiquitously in data systems~\cite{bayer1970organization, comer1979ubiquitous}. The B-tree is 
proven 
to be equivalent to 
a
deterministic skiplist~\cite{munro1992deterministic}. This is a skiplist that can provide a bounded cost even in the worst case. Unlike a probabilistic skiplist, a deterministic skiplist is history-dependent~\cite{munro1992deterministic}.

\begin{figure*}[h]
\centering
    \begin{subfigure}{0.89\textwidth}
    \centering
        \includegraphics[width=\linewidth]{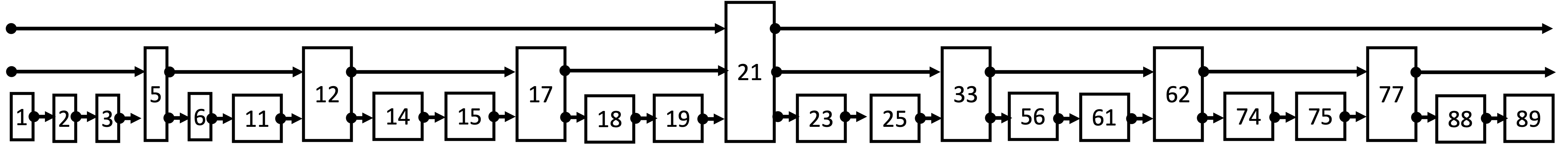}
        \caption{The 1-2-3 skiplist}\label{fig:1-2-3_skiplist}
    \end{subfigure}
    \vfill
    \begin{subfigure}{0.76\textwidth}
    \centering
        \includegraphics[width=\linewidth]{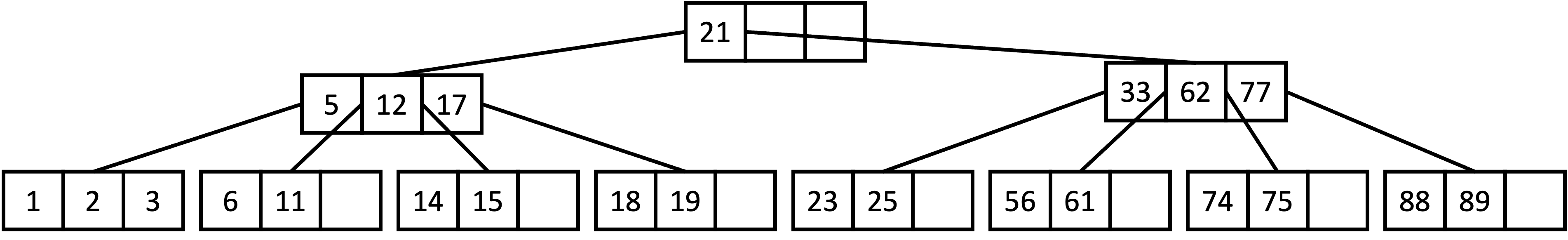}
        \caption{The 2-3-4 tree}\label{fig:2-3-4-tree}
    \end{subfigure}
    \vfill
    \begin{subfigure}{0.76\textwidth}
    \centering
        \includegraphics[width=\linewidth]{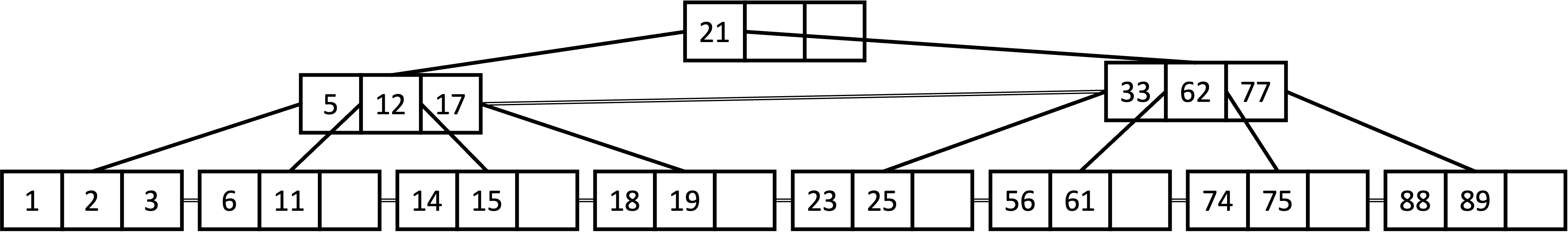}
        \caption{The corresponding Bd-tree}\label{fig:bd-tree}
    \end{subfigure}
    \vfill
    \begin{subfigure}{0.85\textwidth}
    \centering
        \includegraphics[width=\linewidth]{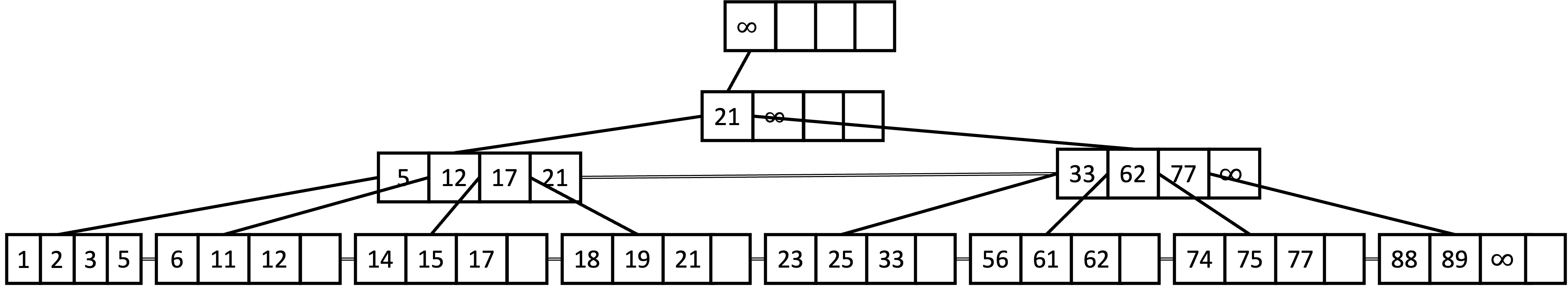}
        \caption{The corresponding Bd$^+$-tree}\label{fig:bdplus-tree}
    \end{subfigure}
\caption{The 1-2-3 skiplist and its counterpart tree structures}\label{fig:skiplist-tree}
\end{figure*}

The 1-2 skiplist and 1-2-3 skiplist~\cite{munro1992deterministic} 
are deterministic variants of the 
skiplist 
in that both their average and worst-case performance are bounded by $\Theta ($lg$n)$. An example 1-2-3 skiplist is given in Figure~\ref{fig:1-2-3_skiplist}. For a 1-2 skiplist of $N$ elements, there exist 1 or 2 elements at Height $h-1$ between any two nodes of Height $h$ ($h>1$). If the forward pointers of a node are stored in an array, when a node grows, a new array has to be allocated with the pointers being copied. This can result in up to $\Theta ($lg$^2n)$ time for one insert~\cite{munro1992deterministic}. Munro et al.~\cite{munro1992deterministic} propose an \textit{array implementation of the 1-2 skiplist} to make arrays have exponentially increasing sizes. During one insert, there is at most one array that needs to be allocated with data being copied~\cite{munro1992deterministic}. The worst-case cost of search, insert, and delete are $\Theta ($lg$n)$~\cite{munro1992deterministic}. These operations are explained in Section~\ref{subsec:deterministic}.

Munro et al.~\cite{munro1992deterministic} show that there is a one-to-one correspondence between the 1-2 skiplist and the 2-3 tree, as well as between the 1-2-3 skiplist in Figure~\ref{fig:1-2-3_skiplist} and the 2-3-4 tree in Figure~\ref{fig:2-3-4-tree}.

The equivalence is formally defined by Lamoureux and Nickerson~\cite{lamoureux1996equivalence} as {\em structural } and {\em functional equivalences}. Structural equivalence requires that data items $A$ and $B$ are logically adjacent to each other in both structures. In contrast, functional equivalence means that the worst-case cost functions are in the same big-$O$ order of complexity. Two intermediate tree structures are defined in~\cite{lamoureux1996equivalence}: The Bd-tree and the Bd$^+$-tree. To construct a Bd-tree, a 2-3-4 tree is used, and each of the tree nodes is connected with its right sibling whenever these sibling nodes exist. Figure~\ref{fig:bd-tree} shows a Bd-tree of order 4, and is equivalent to a B-tree~\cite{bayer1970organization, comer1979ubiquitous} of order 4. The Bd$^+$-tree adds the succeeding key from the parent node as the rightmost entry in each node shown in Figure~\ref{fig:bdplus-tree}.

There are three pair-wise equivalences: B-tree vs. Bd-tree, Bd-tree vs. Bd$^+$-tree, Bd$^+$-tree vs. the horizontal array format of a deterministic skiplist. Lamoureux et al., \cite{lamoureux1996equivalence} prove that a B-tree of order $m$ is structurally and functionally equivalent to the linked list implementation of a deterministic skiplist of order $m$. Also, as in Figure~\ref{fig:bdplus-tree}, the Bd$^+$-tree is similar to the B$^{link}$-tree~\cite{lehman1981blink}, where each node points to its right sibling. However, there are two differences. The Bd$^+$-tree has an extra root node that contains $\infty$. Moreover, each non-leaf node of the Bd$^+$-tree contains an extra key. Showing the equivalence between the B-tree and the skiplist establishes the foundation that the skiplist can be used as a database index in place of a B-tree.

\section{Optimizing Skiplist Design}\label{sec:skiplist-opt}
In this section, we summarize the common optimizations applied to the skiplist in Table~\ref{tab:opt}, and discuss them in detail in this section and their use cases  in Sections~\ref{sec:hardware} to \ref{sec:more-use}.

\begin{table}[h]
  \caption{Summary of Common Optimizations on Skiplists}
  \label{tab:opt}
  \begin{tabular}{cl}
    \toprule
    Common Optimizations&Use Cases\\
    \midrule
    \multirow{3}{*}{Lock-based Concurrent Skiplist} 
     & Basic~\cite{pugh1998concurrent}, Optimistic Lock-based~\cite{herlihy2007simple, herlihy2006provably}, T-List~\cite{mei2017concurrent},\\
     & Unrolled~\cite{platz2019concurrent}, Flat-Combining~\cite{hendler2010flat}, Splay-list~\cite{aksenov2023splay}, \\
     & PhaST~\cite{li2022phast}, ListDB~\cite{kim2022listdb}, GFSL~\cite{moscovici2017gpu}, Leap-List~\cite{avni2013leaplist}\\
    \hline
    \multirow{4}{*}{Lock-free Concurrent Skiplist} 
     &Two-Step Deletion~\cite{sundell2004lockfree}, Three-Step Deletion~\cite{fomitchev2004lock}, KiWi~\cite{basin2020kiwi}, \\
     & Fraser's Design~\cite{fraser2004practical}, No Hotspot~\cite{crain2013no}, Nitro~\cite{lakshman2016nitro},\\
     & Java Implementation~\cite{openjdkmirror}, Rotating~\cite{dick2017rotating}, PI/PSL~\cite{xie2016pi,xie2017parallelizing},\\
     & UPSkiplist~\cite{chowdhury2021scalable}, CMSL~\cite{gpulockfree}, SprayList~\cite{alistarh2015spraylist}\\
    \hline
    MVCC Skiplist 
     &KiWi~\cite{basin2020kiwi}, Nitro~\cite{lakshman2016nitro}, JellyFish~\cite{yeon2020jellyfish}, Jiffy~\cite{kobus2022jiffy}, Frugal~\cite{kim2021frugal}\\
    \hline
    \multirow{3}{*}{Deterministic Skiplist}
     &Deterministic~\cite{munro1992deterministic}, No Hotspot~\cite{crain2013no},\\
     & $k$-d Skiplist~\cite{nickerson1998skip}, $k$-d Range Skiplist~\cite{lamoureux2005deterministic},\\
     &Network Overlay algorithms~\cite{clouser2008tiara, nor2013corona,mandal2012deterministic,singh2015concurrent}\\
    \hline
    \multirow{3}{*}{Other Promotion Heuristics}
     & Biased~\cite{ergun2001biased}, ($a, b$)-Biased And Randomized Biased~\cite{bagchi2005biased},\\
     & Self-Adjusting~\cite{ciriani2002static, ciriani2007data}, T-List~\cite{mei2017concurrent}, Splay-list~\cite{aksenov2023splay}, \\
     & NV-skiplist~\cite{chen2019design}\\
    \hline
    \multirow{2}{*}{Partitioned Skiplist} 
     &NV-skiplist~\cite{chen2019design}, PhaST~\cite{li2022phast}, ListDB~\cite{kim2022listdb}, PI/PSL~\cite{xie2016pi,xie2017parallelizing}, \\
     & RS-store~\cite{huang2021rs}\\
    \hline
    \multirow{4}{*}{Unrolled Node}
     &KiWi~\cite{basin2020kiwi}, Cache-sensitive, ~\cite{sprenger2017cache}, Write-optimized~\cite{bender2017write}, \\
     &FlashSkiplist~\cite{wang2017flashskiplist}, NV-skiplist~\cite{chen2019design}, PhaST~\cite{li2022phast}, GFSL~\cite{moscovici2017gpu}, \\
     & CMSL~\cite{gpulockfree}, NMP-skiplist~\cite{liu2017nmp,choe2019nmp}, HybriDS~\cite{choe2022hybrids}, S3~\cite{zhang2019s3},\\
     & Leap-List~\cite{avni2013leaplist}\\
  \bottomrule
\end{tabular}
\end{table}

\subsection{Concurrent Skiplists}\label{subsec:concurrent-skiplists}
Concurrent access and operations to the skiplist can be supported by either lock or lock-free techniques. Both implementations are possible with linear-scalability~\cite{pugh1998concurrent}. Also, there are variants of skiplists that support multi-versioning and transactional memory~\cite{yeon2020jellyfish, kobus2022jiffy, huang2019x, kardarasfast}. In the following subsections, we discuss these techniques in further detail.

\subsubsection{The Lock-based Skiplist}\label{subsubsec:lock-based}
\hfill

\noindent
{\bf The Basic Skiplist~\cite{pugh1998concurrent}.}
Pugh~\cite{pugh1998concurrent} uses locks for concurrent access and manipulation of skiplists. Each node maintains an array of pointers at each level the node spans. The search procedure is the same as the basic skiplist search algorithm. Insertion starts by first searching the inserted key and finding the node that contains the largest key that is smaller than the inserted key on the bottom level. After the node \textit{prev} (the node after which the new node is to be inserted) is found, \textit{prev} is locked and a new node is created, and is inserted after it in bottom level. Next, the height of the node is determined and the pointers in upper levels are updated one at a time.
This step is repeated until the desired height.
Similarly, for the delete procedure, the height of the node is decremented to zero, and the steps are the same steps as those for insert. We  term this skiplist  \texttt{BasicSL} (Section~\ref{subsubsec:cc-perf-cmp}).

\noindent
{\bf The Optimistic Lock-based Skiplist~\cite{herlihy2007simple, herlihy2006provably}.}
Herlihy et al.~\cite{herlihy2007simple, herlihy2006provably} have proposed an optimistic lock-based skiplist. It is optimistic because all searches are performed without locks. Only when the item is found, the node and its predecessor are locked, and a validation step ensuring the list has not been changed is performed. Each node maintains two flags, \texttt{marked} that indicates whether a logical deletion is performed or not, and \texttt{fullyLinked} that hints on whether the locked node has been added to all the levels it should reside. During insertion, if the key is not found, all the predecessors are locked up to the height of the new node $H$. Then, the procedure validates that the predecessors are adjacent with the successors up to $H$, and they are not marked for deletion. Next, the new node is created, and is inserted into the structure. Finally, \texttt{fullyLinked} is set when the insert procedure is performed, and all locks are released. We  term this skiplist  \texttt{LazySL}  (Section~\ref{subsubsec:cc-perf-cmp}).

\noindent
{\bf The Unrolled Skiplist~\cite{platz2019concurrent}.}
The unrolled skiplist~\cite{platz2019concurrent} uses \textit{group mutual exclusion} as its locking technique. It groups multiple keys per node to achieve a better cache locality as well as reduce the number of pointer chases. As multiple keys share the same node, multiple threads performing the same operation on different keys in the same node can proceed concurrently. In order for a thread to join a session, \cite{platz2019concurrent} uses two algorithms. One algorithm uses a word to track the type of the session and the number of threads in the session. When a thread wants to join, it repeatedly reads the word and updates the word via compare-and-swap (CAS) if it joins the session successfully. The second algorithm uses an exclusive lock to protect. The threads push themselves into a queue and wait until session is compatible. We term this skiplist  \texttt{UnrolledSL} (Section~\ref{subsubsec:cc-perf-cmp}).

\noindent

{\bf The Flat-Combining Skiplist~\cite{hendler2010flat}.}
Flat-combining is a synchronization technique based on coarse locking~\cite{hendler2010flat}. 
In this approach, the skiplist is protected by a shared lock.
A publication list, sized proportionally to the number of threads, is maintained. 
When a thread attempts to access the skiplist, it adds a record to the publication list, specifying the operation it needs to perform. The threads  compete for the shared lock.
If a thread acquires the lock, it becomes the combiner. 
The combiner scans the publication list and applies the \textit{combined requests} from all threads to the skiplist. 
Threads that fail to acquire the lock wait for the combiner to execute their requests, and return the results in the respective request fields~\cite{hendler2010flat}.

\subsubsection{The Lock-free Skiplist}\label{subsubsec:lock-free}
\hfill
\begin{figure*}[h]
\centering
    \begin{subfigure}{0.17\textwidth}
    \centering
        \includegraphics[width=\linewidth]{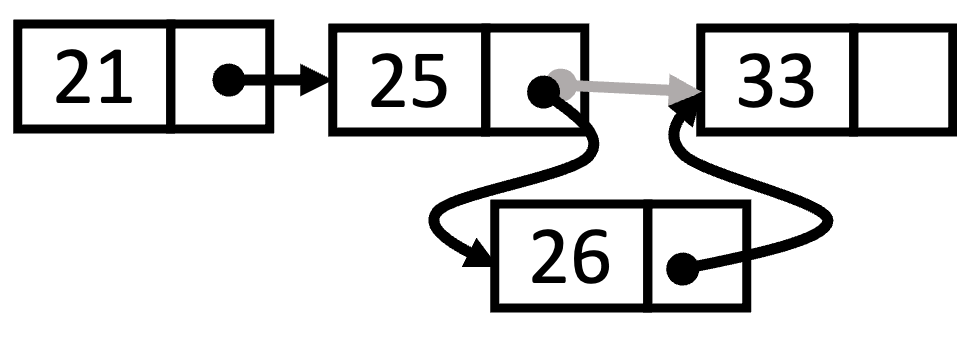}
        \caption{}
        \label{fig:ll-insert}
    \end{subfigure}
    \hfill
    \begin{subfigure}{0.17\textwidth}
    \centering
        \includegraphics[width=\linewidth]{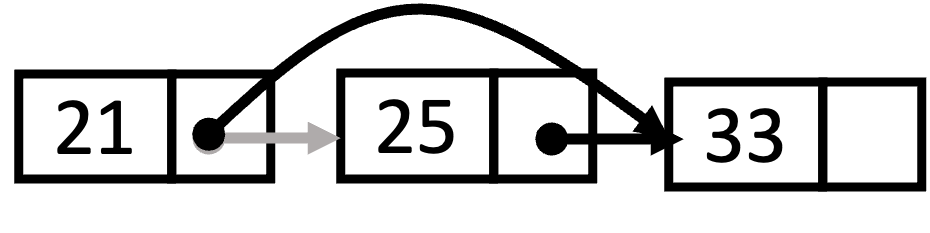}
        \caption{}
        \label{fig:ll-delete-old}
    \end{subfigure}
    \hfill
    \begin{subfigure}{0.17\textwidth}
    \centering
        \includegraphics[width=\linewidth]{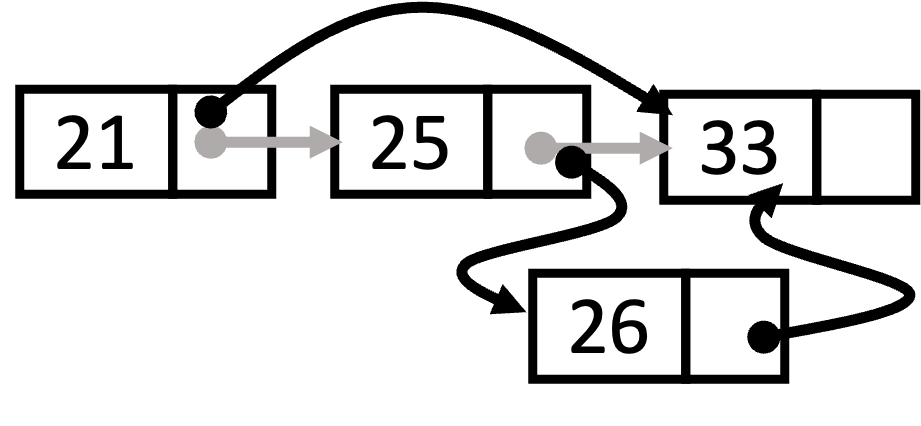}
        \caption{}
        \label{fig:ll-delete-dilemma}
    \end{subfigure}
    \hfill
    \begin{subfigure}{0.17\textwidth}
    \centering
        \includegraphics[width=\linewidth]{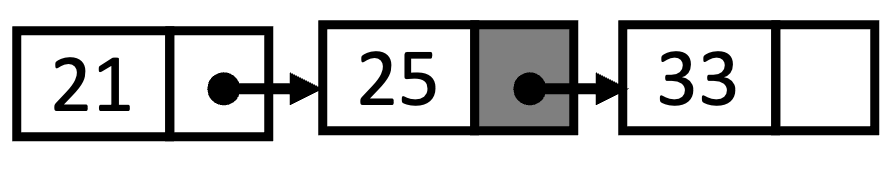}
        \caption{}
        \label{fig:ll-delete-1}
    \end{subfigure}
    \hfill
    \begin{subfigure}{0.17\textwidth}
    \centering
        \includegraphics[width=\linewidth]{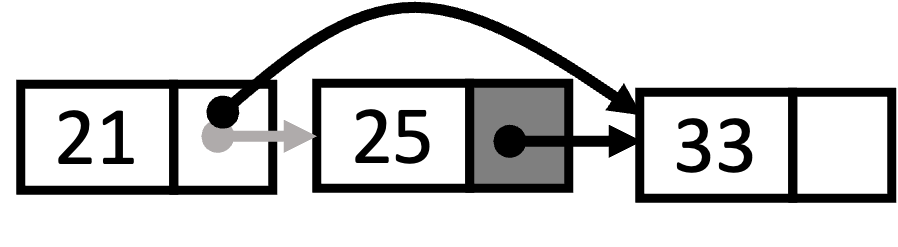}
        \caption{}
        \label{fig:ll-delete-2}
    \end{subfigure}
\caption{The lock-free linked list with two-step deletion. (a) A single CAS operation insertion. (b) A single CAS operation deletion. (c) A lost update. (d) Step 1 of the two-step deletion. (e) Step 2 of the two-step deletion}\label{fig:linked-list}
\end{figure*}

\noindent
{\bf The Lock-Free Linked List~\cite{harris2001pragmatic}.}
Harris~\cite{harris2001pragmatic} introduces a lock-free implementation of linked lists via the atomic operation compare-and-swap (CAS). In Figure~\ref{fig:ll-insert}, to insert a new node, say Node 26, the next pointer of Node 25 is atomically pointed to the new node. However,  deleting Node 25 with one CAS as  in Figure~\ref{fig:ll-delete-old} may result in a situation as in Figure~\ref{fig:ll-delete-dilemma} where inserting Node 26 is lost because the insert finds Node 25 before Node 25 is deleted. A two-step delete can avoid this situation by marking the next pointer of Node 25 as invalid with one CAS instruction, that is referred to as a logical delete. The second step is a physical deletion that changes the pointer atomically. The design of the lock-free linked list can be applied to the skiplist~\cite{fraser2004practical}.

\noindent
{\bf The Lock-free Linked List with Two-Step Deletion~\cite{harris2001pragmatic}.}
Harris~\cite{harris2001pragmatic}  uses a two-step process to delete a node. In the first step, the node's next pointer is marked to indicate that the node is logically deleted.  A pointer can be marked by setting to 1 the lowest bit in the pointer field, as the two lowest order bits are set to 0 in a 32-bit system.  In the second step, the node is physically deleted.
As in Figure~\ref{fig:ll-delete-1}, observe that  Node 25's next pointer is marked in the first step, making it logically deleted. Assume that Node 26 is being concurrently inserted, now it identifies  Node 25 as logically deleted during traversal and  restarts the insert process from the beginning until Node 25 is physically deleted. This  avoids the problem of disappearing nodes discussed above. 

\noindent
{\bf The Lock-free Skiplist with Two-Step Deletion~\cite{sundell2004lockfree}.
Since skiplists are ordered and provide a delete operation, they can be used to create a priority queue by deleting the first (smallest) element. Sundell and Tsigas create a lock-free skiplist~\cite{sundell2004lockfree} as well as a concurrent priority queue (ref Section~\ref{subsec:priority-queue})~\cite{sundell2005fast} based on the two-step delete and the CAS primitive.}

\begin{figure*}[h]
\begin{minipage}{.2\linewidth}
\centering
\begin{subfigure}{\linewidth}
    \includegraphics[width=\textwidth]{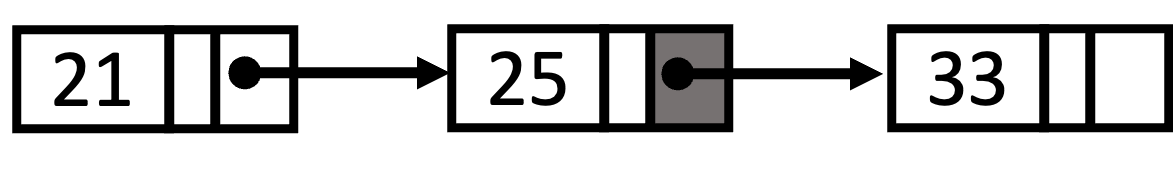}
    \caption{}\label{fig:ll-3step-1}
\end{subfigure} \\
\begin{subfigure}{\linewidth}
    \includegraphics[width=\textwidth]{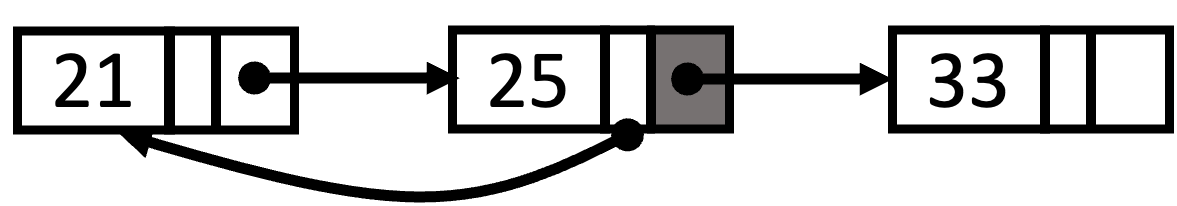}
    \caption{}\label{fig:ll-3step-2}
\end{subfigure} \\
\begin{subfigure}{\linewidth}
    \includegraphics[width=\textwidth]{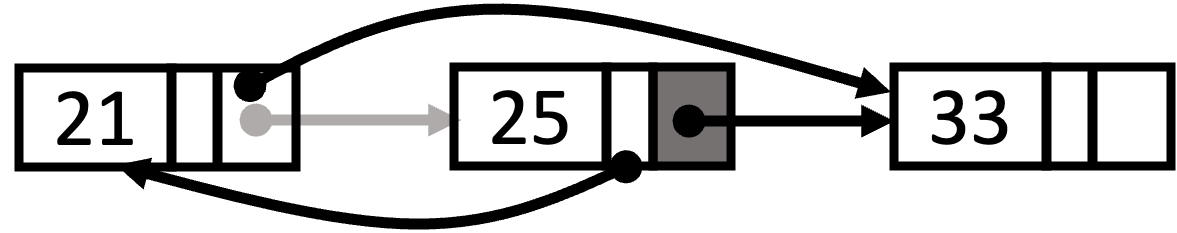}
    \caption{}\label{fig:ll-3step-3}
\end{subfigure} 
\end{minipage}
\begin{minipage}{.5\linewidth}
\centering
\begin{subfigure}{.9\linewidth}
    \includegraphics[width=\textwidth]{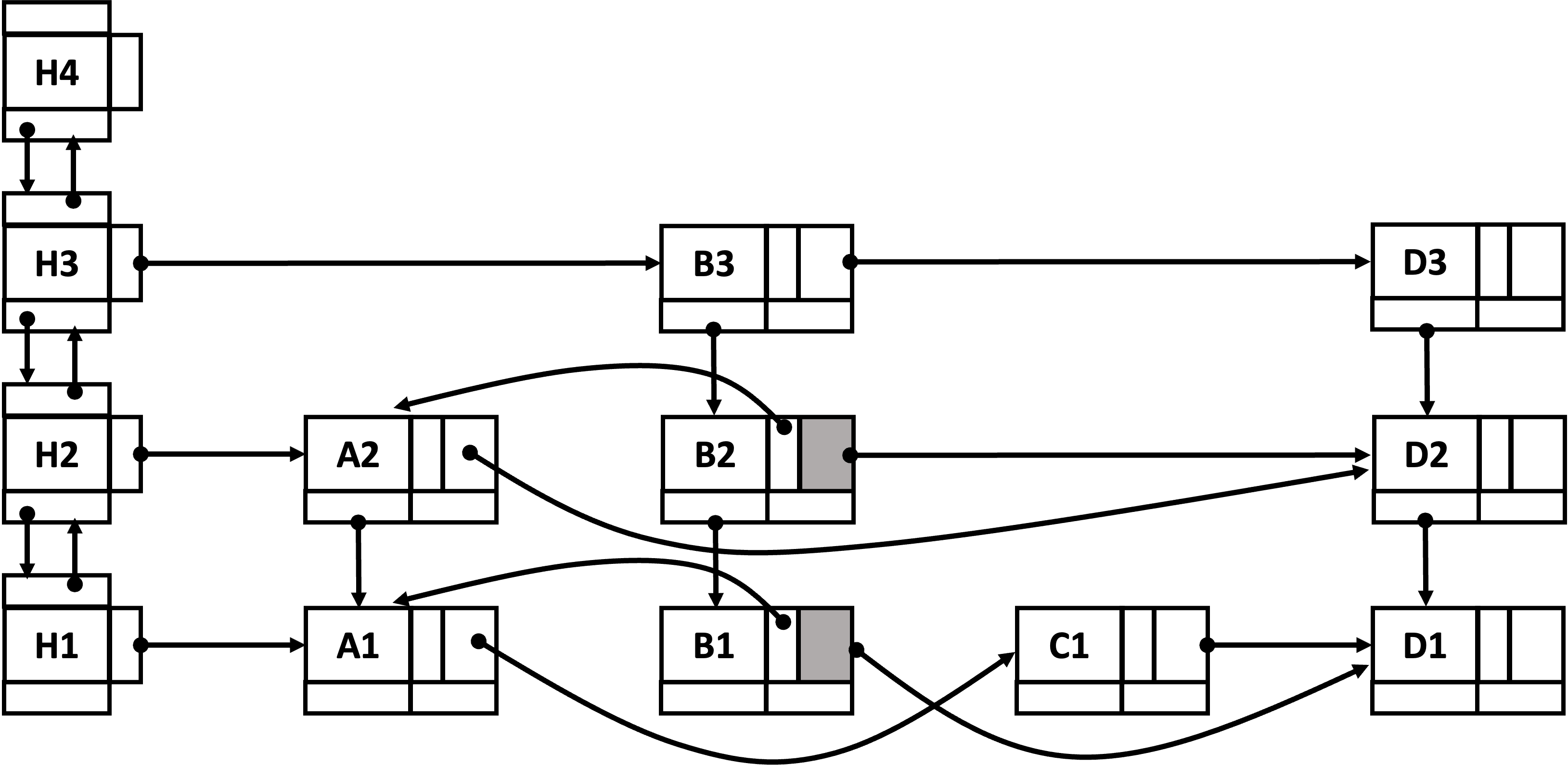}
    \caption{}\label{fig:fomichev_skiplist}
\end{subfigure}
\end{minipage}
\caption{The 3-step deletion in a lock-free linked list ((a) to (c)) and (d) the lock-free skiplist with Three-Step Deletion~\cite{fomitchev2004lock}.}\label{fig:ll-3step}
\end{figure*}


\noindent
{\bf The Lock-free Linked List with Three-Step Deletion~\cite{fomitchev2004lock}.}
The two-step deletion in~\cite{harris2001pragmatic} restarts the insert process when there is a conflict. 
Thus, the average cost  is increased~\cite{fomitchev2004lock}. Fomitchev et al.~\cite{fomitchev2004lock} use a three-step deletion process to avoid starting a failed CAS from the start of the list. In the first step, Node 25's next pointer is marked (Figure~\ref{fig:ll-3step-1}). Next, a back pointer is created to point to the previous node, say Node 21 (Figure~\ref{fig:ll-3step-2}). This is performed to prevent restarting the process from the beginning of the list. In the final step, Node 25 is physically deleted  (Figure~\ref{fig:ll-3step-3}). By following this three-step deletion, When Node 26 is being inserted, it can start the traversal from Node 21 (with the help of the back pointer) instead of starting from the beginning of the list.

\noindent
{\bf The Lock-free Skiplist with Three-Step Deletion~\cite{fomitchev2004lock}.}
Fomitchev and Ruppert~\cite{fomitchev2004lock} implement a lock-free skiplist that is composed of the above linked list. Each skiplist node that has $k$ forward pointers is divided into $k$ nodes vertically. Only the bottom one (\textit{root node}) contains the data. These $k$ nodes form a linked list vertically, and is called a {\em tower}. For example, in Figure~\ref{fig:fomichev_skiplist}, $B1$, $B2$ and $B3$ form a tower. Both $B2$ and $B3$ have a pointer to $B1$ where data is stored. Horizontally, a back pointer field is added to each node to facilitate deletion. Insert starts by finding the right location for the new data, and inserts the \textit{root node} first. Other tower nodes can be added bottom up. Delete first marks the \textit{root node}, and utilizes the back pointers in each level as in the above lock-free linked list. When the \textit{root node} is marked deleted, the tower nodes are superfluous. $B2$ and $B3$ are superfluous. Search in the skiplist helps delete superfluous nodes.

\noindent
{\bf Fraser's Design~\cite{fraser2004practical}.}
In~\cite{fraser2004practical}, several designs of a lock-free skiplist are included. A compare-and-swap (CAS)-based design uses the idea of the lock-free linked list~\cite{harris2001pragmatic} as each level in the skiplist is an independent linked list. There can be inconsistencies in concurrent deletes and inserts using only one CAS as illustrated in Figure~\ref{fig:linked-list}. Thus, a marking scheme~\cite{harris2001pragmatic} is adopted~\cite{fraser2004practical}. The second lock-free skiplist~\cite{fraser2004practical} uses multi-word CAS (MCAS). For inserts, all writes are grouped in a batch, and are installed via a single MCAS operation, while reads check that the read location is not owned by the MCAS operation. The third lock-free technique uses transactional memory~\cite{harris2022transactional}. Each node is treated as a separate transactional object. Concurrent operations on the same node are performed sequentially within a transaction. We term this skiplist  \texttt{FraserSL} (Section~\ref{subsubsec:cc-perf-cmp}).

\begin{figure}[h]
\centering
    \begin{subfigure}{0.47\textwidth}
    \centering
        \includegraphics[width=\linewidth]{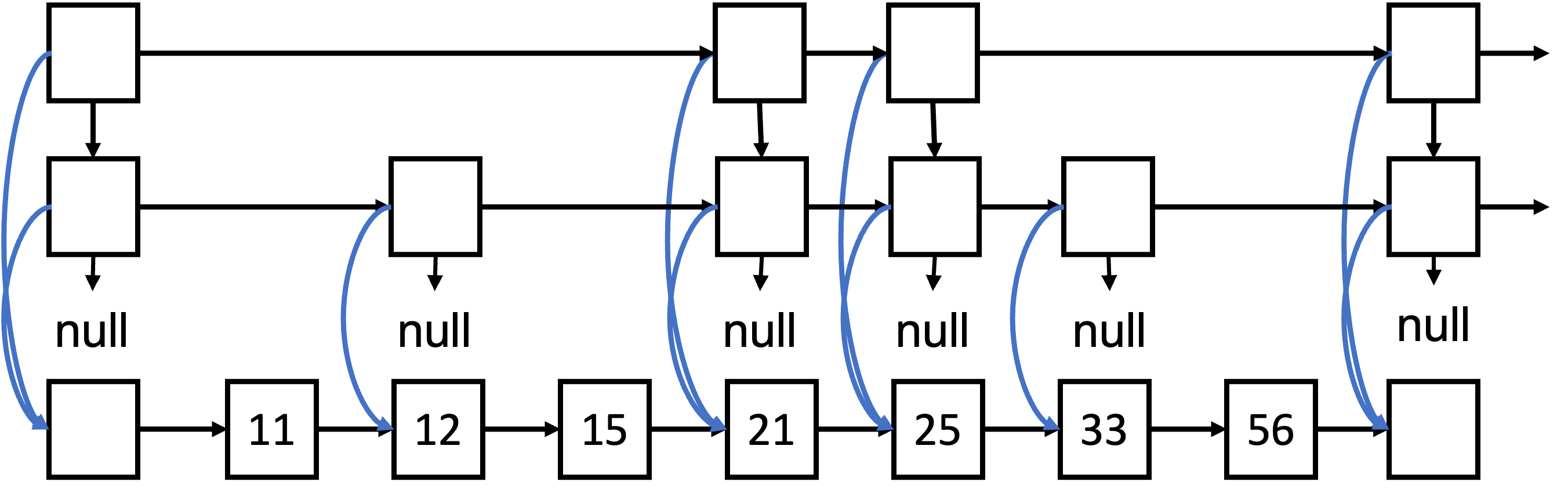}
        \caption{}
    \label{fig:java_skiplist}
    \end{subfigure}
    \hfill
    \begin{subfigure}{0.47\textwidth}
    \centering
        \includegraphics[width=\linewidth]{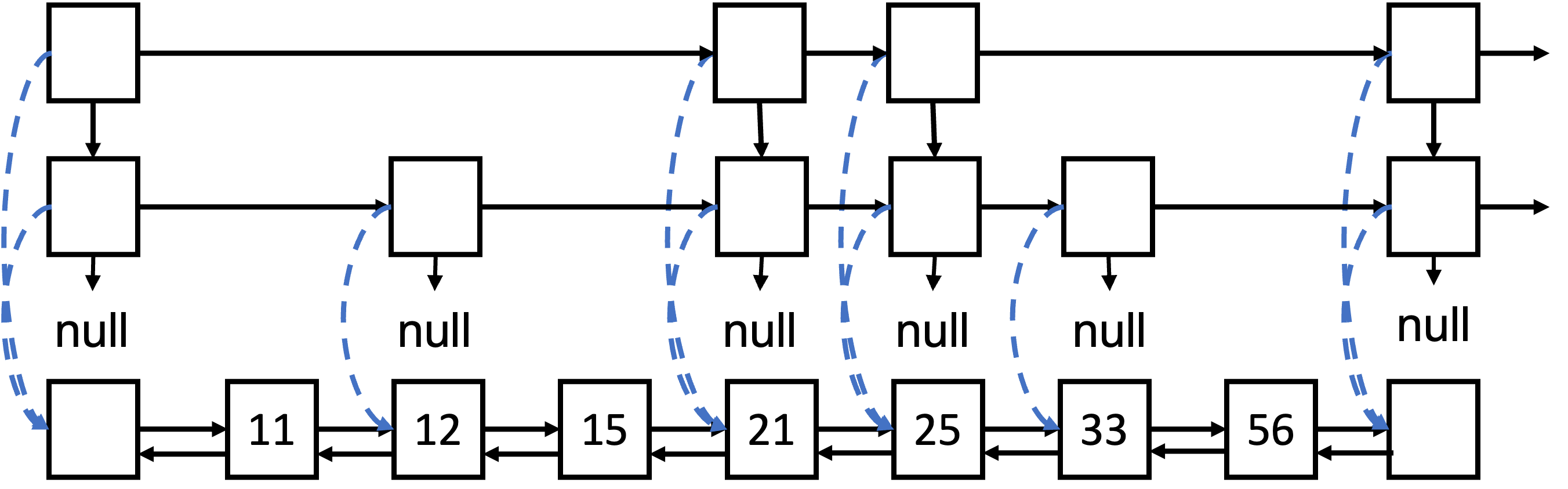}
        \caption{}
    \label{fig:no_hotspot}
    \end{subfigure}
    \vfill
    \begin{subfigure}{\textwidth}
    \centering
        \includegraphics[width=0.78\linewidth]{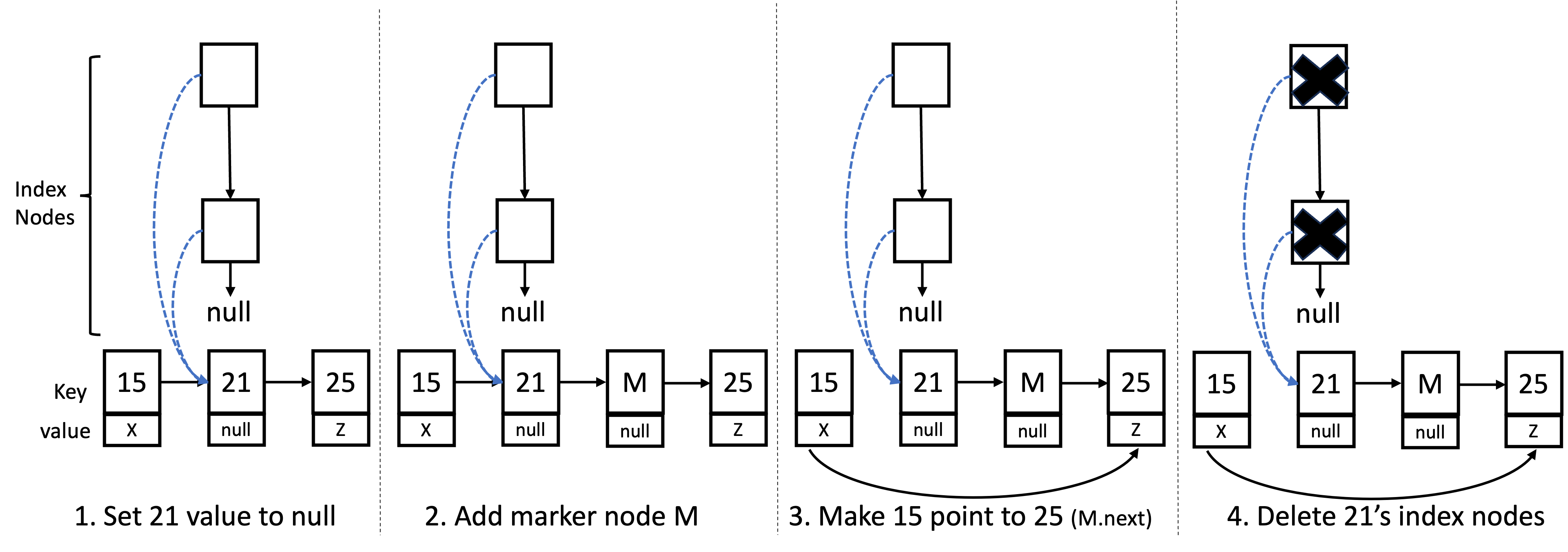}
        \caption{}
    \label{fig:java_skiplist_deletion}
    \end{subfigure}
    \caption{(a) Concurrent skiplist Java implementation~\cite{openjdkmirror}, (b) the no hotspot skiplist~\cite{crain2013no}, (c) deletion of Node 21 from (a) }\label{fig:java}
\end{figure}

\noindent
{\bf Concurrent Skiplist Java Implementation~\cite{openjdkmirror}.}
\texttt{ConcurrentSkipListMap}, implemented by Doug Lea~\cite{openjdkmirror}, is part of the \texttt{java.util.concurrent} library. It relies on Harris's,  Fomitchev and Ruppert's, and Fraser's approaches~\cite{harris2001pragmatic, fomitchev2004lock, fraser2004practical}. Fraser's implementation uses Compare-And-Swap (CAS)-able node pointers. When a node is to be deleted, it is replaced or spliced with another node that represents a delete mark. This technique reduces space overhead by not using delete marks, which improves performance as well.
A node is logically deleted when the value is null or when its next node is marked, but is physically deleted when the garbage collector collects it. A garbage collector collects a deleted node during traversal when it finds a node that is marked as logically  deleted (i.e., when the node's value is null). The key idea  is to use Compare And Swap (CAS)  to detect an insert or delete in the data structure. 
If  CAS  fails, then a concurrent thread  has inserted or deleted a node at the  same time. 
Thus, the operations are restarted.
The search operation reports \texttt{Not Found} if a logically deleted node is encountered.
Insert proceeds by first searching for the predecessor of the inserted value. The predecessor is recorded in case of a concurrent insert or delete.
Once  CAS succeeds, the new node is inserted into the data layer. 
Then, the height is determined by Algorithm~\ref{alg:random-level}, and the index nodes are inserted in the upper levels via CAS  to ensure correctness. 
Figure~\ref{fig:java_skiplist_deletion} gives an example of deleting Node 21.
In the figure, for simplicity, the index nodes for only Node 21 are displayed. 
First, the value field is set to null atomically (Figure~\ref{fig:java_skiplist_deletion}-1). Next, the next pointer of Node 21 is set atomically to point to a new marker node $M$ to avoid the modification of the next pointer of Node 21. 
(Figure~\ref{fig:java_skiplist_deletion}-2). 
Then, atomically switch the predecessor of Node 21 to point to the next node of Node $M$.
(Figure~\ref{fig:java_skiplist_deletion}-3).
Finally, the index nodes of Node 21 are deleted. Node 21 in the bottom level is garbage collected after all of its index nodes are deleted (Figure~\ref{fig:java_skiplist_deletion}-4). We term this skiplist  \texttt{JavaImpl}  (Section~\ref{subsubsec:cc-perf-cmp}).

\noindent
{\bf The No Hotspot Skiplist~\cite{crain2013no}.}
Nodes at the higher levels of a skiplist are accessed frequently as all the queries start from the highest level. There is thread contention at the higher levels, which leads to poor performance. Thread contention primarily occurs due to inserts.
Contention Friendly Skiplist~\cite{crain2013no}  avoids this issue by decoupling inserts at the data layer from those at the index layers, similarly for deletes. It uses a deterministic skiplist to balance the nodes. Also, it uses a background thread, termed the ``adaptive thread'', to have the changes in the data layer propagate to the index layers, including balancing by promoting or demoting the level of the node, lowering towers, and garbage collecting the locally deleted nodes. 
Another key change is the use of a doubly linked node in the data layer, as shown in Figure~\ref{fig:no_hotspot}. Doubly linked nodes are needed to prevent the repeated find predecessor function calls that can be performed in only $O(1)$ time when using doubly-linked nodes compared to $O(\log n)$ time for the earlier approach. 
Search is performed identically to the Java implementation (Section~\ref{subsubsec:lock-free}), except that the previous pointer is used instead of invoking the \texttt{findPredecessesor} function when CAS fails. 
Inserts are first performed at the data layer. The adaptive thread promotes the level of the node to maintain balance. 
Deletes take place at the data layer in a way similar to Java's implementation. Since deletes have been decoupled in the index level, this is handled by the adaptive thread. We term this skiplist  \texttt{CFSL} (Section~\ref{subsubsec:cc-perf-cmp}).

\begin{figure*}[h]
\centering
        \includegraphics[width=0.6\linewidth]{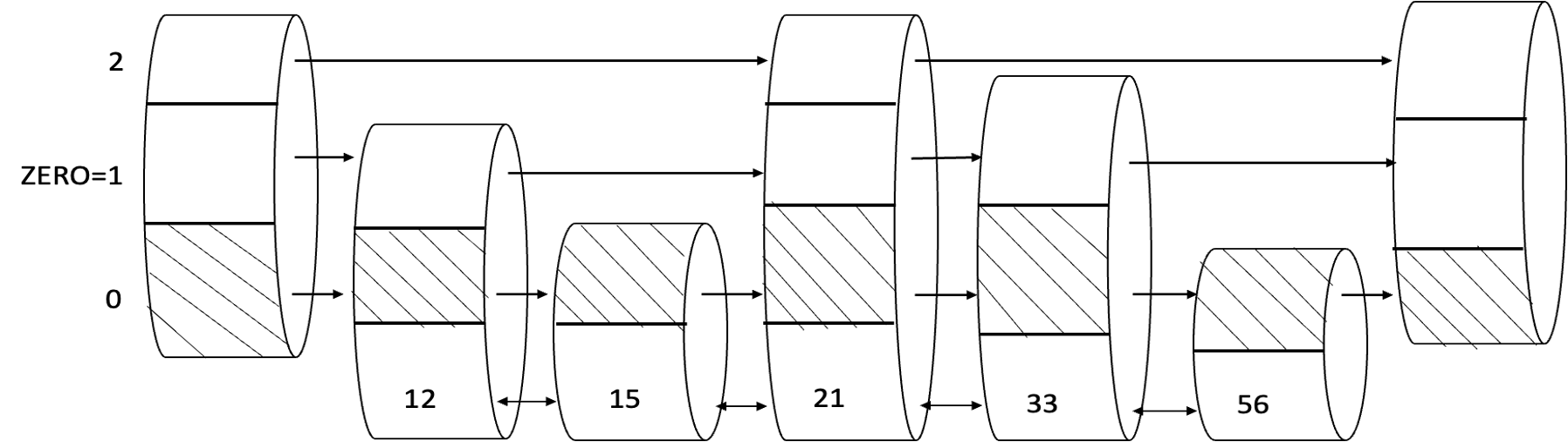}
\caption{The Rotating Skiplist and lowering towers in the rotating skiplist~\cite{dick2017rotating}.}\label{fig:rotating}
\end{figure*}

\noindent
{\bf The Rotating Skiplist~\cite{dick2017rotating}.}
Although the no-hotspot skiplist avoids contention in the upper levels and remains lock-free, its performance does not scale well with the increase in the size of the number of elements due to high cache misses~\cite{dick2017rotating}. The no-hotspot skiplist uses linked nodes, and has poor spatial locality. It also requires linear time to lower the level of the skiplist during rebalancing. The {\em Rotating Skiplist}~\cite{dick2017rotating} addresses these issues by using a {\em wheel} and a global variable named ``ZERO''. ZERO is used to indicate the lowest logical level of all the arrays
and
ZERO initially points to level 0. A wheel is an array node, where by using an array, spatial locality and cache hits are improved. The Rotating Skiplist uses modular arithmetic to wrap around the array node (level = (current level + ZERO) \texttt{mod} max-levels).  The Rotating skiplist has wheels that  ``rotate" during the lowering of the skiplist to balance the structure, and hence providing good time complexity for all the operations. The Rotating skiplist contains a doubly-linked list in the bottom-most level, similar to~\cite{crain2013no},  where this level contains the actual data. All the operations except Lowering Towers are performed in a similar manner to that of a no-contention skiplist. 

Lowering Towers is performed by one atomic instruction by incrementing the ZERO variable by one. As in Figure~\ref{fig:rotating}, ZERO is changed from 0 to 1. All operations that are operating concurrently to lowering towers will continue to operate normally. However, any new operation will now ignore Level 0, and operate on Levels 1 and 2 only. After all operations operating on Level 0 are completed, the garbage collector physically deletes Level 0.
The Rotating Skiplist overcomes several problems, e.g., having low-cache miss ratio, scales well with the increase in size of the skiplist, and also lowers the level of the skiplist in one atomic instruction. However, the main disadvantage is that the space complexity of the Rotating Skiplist is $O(n\log n)$~\cite{dick2017rotating}. This space is caused due to the fixed wheel capacity that is predetermined at the start time for each wheel~\cite{dick2017rotating}. A rotating skiplist is used to mitigate contention hotspots, and reduce low cache misses, and thus enables the rotating skiplist to achieve peak performance of 200 MOPS/sec, and scales well with the increase in size of the skiplist~\cite{dick2017rotating}. We term this skiplist \texttt{RotateSL} (Section~\ref{subsubsec:cc-perf-cmp}).

\subsubsection{The Multi-Versioned Skiplist}\label{subsubsec:mvcc}
Multiversion Concurrency Control (MVCC) is an optimistic concurrency control technique that keeps a copy of each data item that is being modified. Each user sees a {\em snapshot} of the database at the time when the user's transaction  starts.  Snapshot isolation~\cite{berenson1995critique} guarantees that a transaction, say $T$, observes a database that is produced by all the transactions that are committed before $T$ starts. Studies compare MVCC algorithms, e.g.,~\cite{bernstein1983multiversion, reed1978naming}.

\noindent
{\bf KiWi~\cite{basin2020kiwi}.} KiWi is a key-value map  built on top of a lock-free skiplist that employs MVCC. KiWi supports long atomic range searches. Version number is managed by range searches that are less frequent. Write  may overwrite the data without updating the version number.

\noindent
{\bf Nitro~\cite{lakshman2016nitro}.} The Nitro storage engine uses a latch-free skiplist as its core index. It supports MVCC. Multiple versions of a key can co-exist in the skiplist with the latest version  considered as a higher key. Deletes in Nitro logically mark the data as deleted. To avoid searching obsolete data, a dedicated thread responsible for garbage collection periodically removes obsolete nodes~\cite{lakshman2016nitro}.

\begin{figure}
    \centering
    \begin{minipage}{.33\linewidth}
            \begin{subfigure}[t]{.9\linewidth}
                \includegraphics[width=\textwidth]{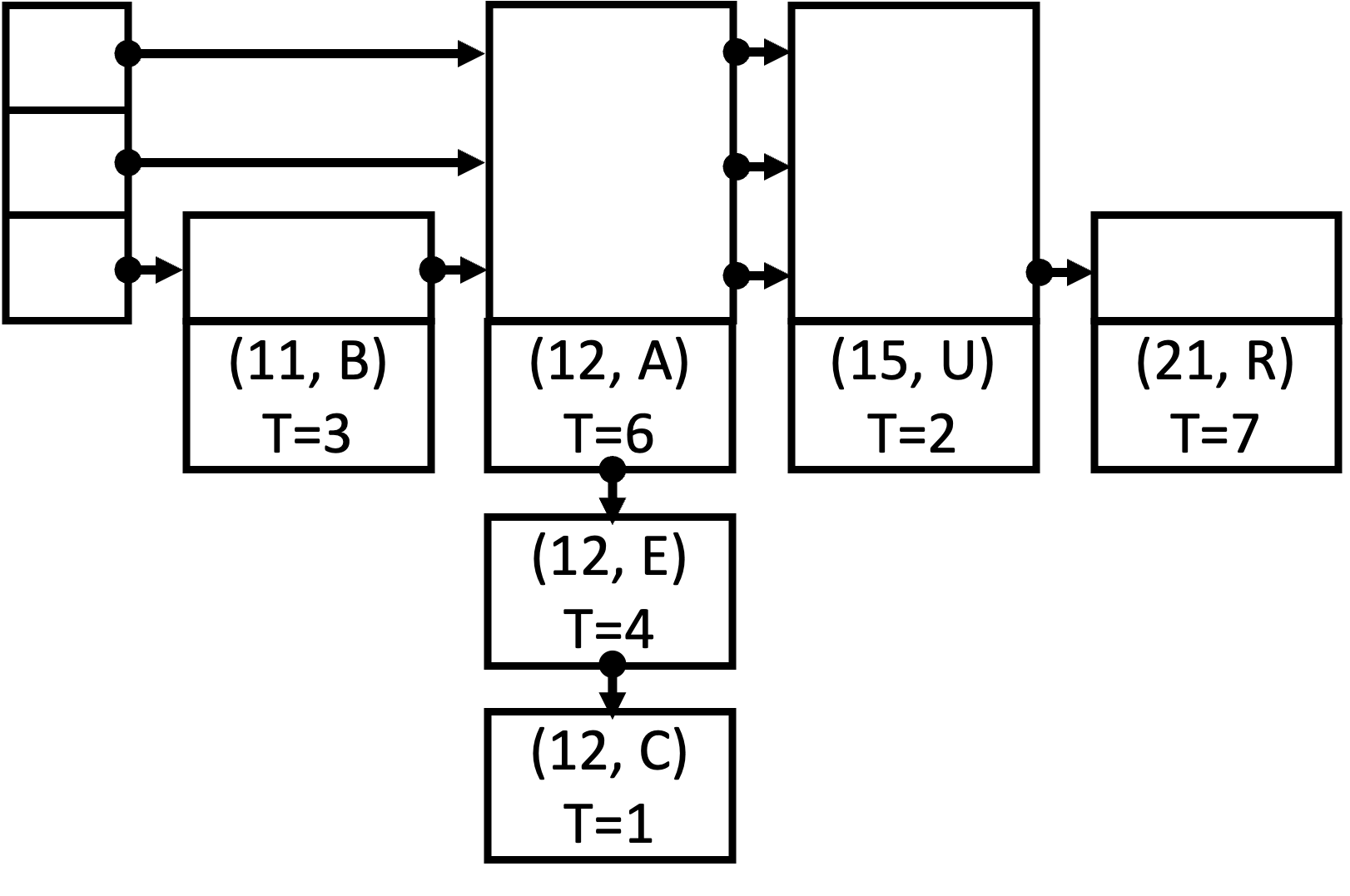}
                \caption{Jellyfish skiplist}\label{fig:jellyfish}
            \end{subfigure}
        \end{minipage}
    \begin{minipage}{.45\linewidth}
        \begin{subfigure}[t]{.9\linewidth}
            \includegraphics[width=\textwidth]{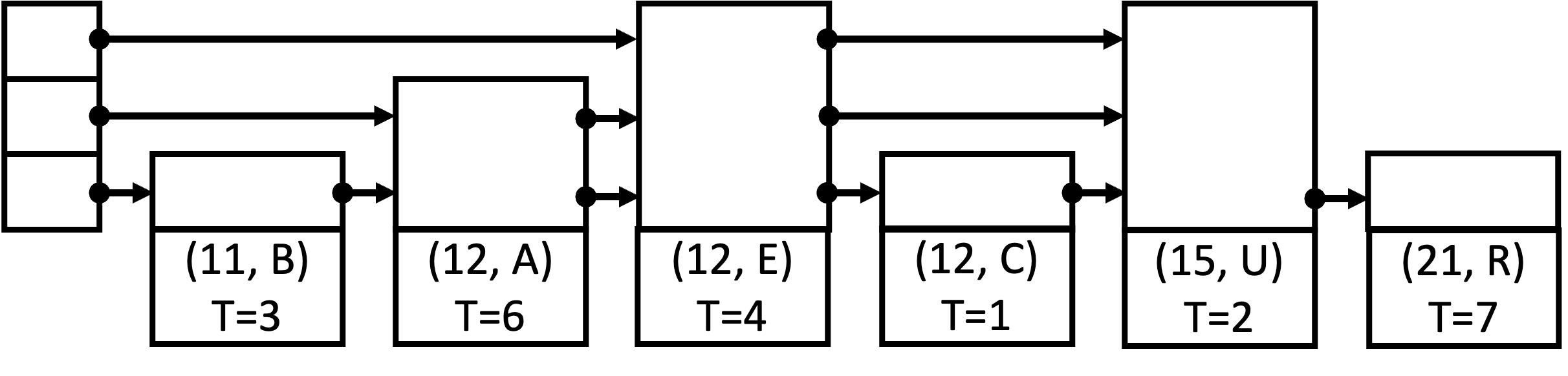}
            \caption{Skiplist that does not consider timestamps.}\label{fig:mvcc}
        \end{subfigure} \\
        \begin{subfigure}[b]{.95\linewidth}
            \includegraphics[width=\textwidth]{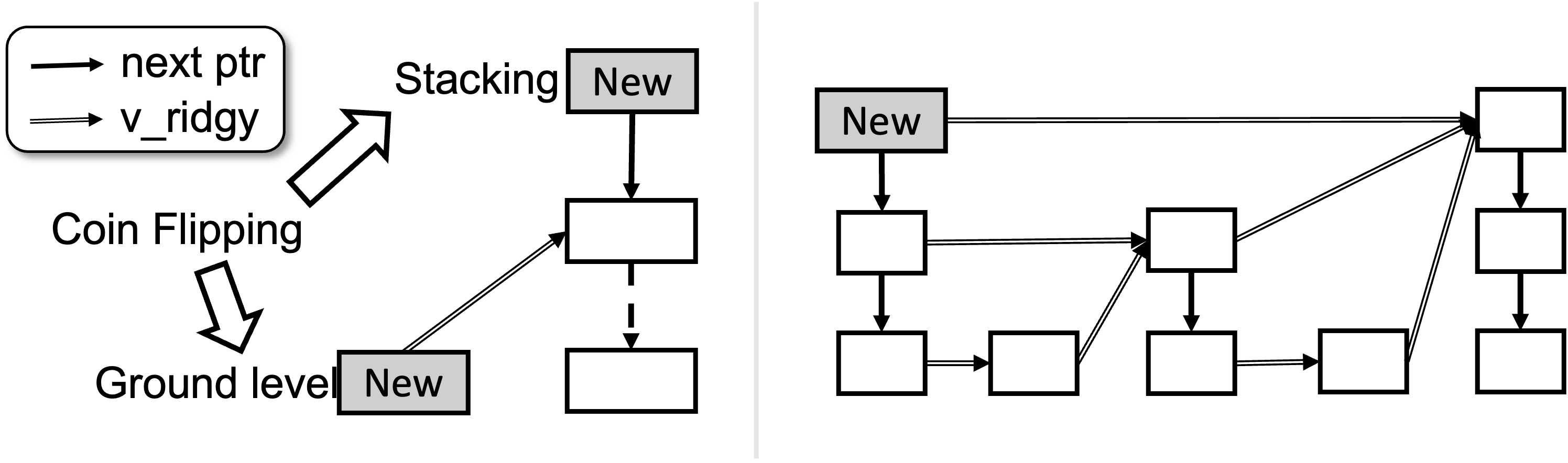}
            \caption{Frugal skiplist}\label{fig:frugal}
        \end{subfigure} 
    \end{minipage}
    \caption{Multi-versioned skiplists.}\label{fig:mvcc-skiplist}
\end{figure}

\noindent
{\bf JellyFish~\cite{yeon2020jellyfish}.}
Both the skiplist of X-Engine~\cite{huang2019x} and JellyFish~\cite{yeon2020jellyfish} are designed with the observation that multiple versions of the same key are linked according to their timestamps at the skiplist's bottom layer. Searching for a key within certain timestamp may require to traverse the list in reverse order if the same key at the upper layer has an obsolete timestamp. In Figure~\ref{fig:mvcc}, searching for Key 12  lands at Value E with Timestamp 4 that is obsolete because of a more recent value at Timestamp 6. Both X-Engine's skiplist and JellyFish organize a vertical list to store multiple versions of the same key (Figure~\ref{fig:jellyfish}). Inserting a new key follows the same process as that of the ordinary skiplist. However, for updates, JellyFish adds the newer version into the vertical list with the most recent version at the second place of the list (The very first head of the vertical list is part of the horizontal bottom layer and is not moved). The nodes for Key 12 are arranged in  timestamp order (Figure~\ref{fig:jellyfish}).Operation {\em Get} starts at the skiplist's upper layer, and searches the target data until it hits the vertical list at the bottom layer. Operation {\em Scan} traverses the bottom layer, and does not have to scan through the obsolete key-value pairs. The read operations, both {\em Get} and {\em Range Search}, benefit the most from this design. We term this skiplist  \texttt{JellyFish} (Section~\ref{subsubsec:cc-perf-cmp}).


\noindent
{\bf Jiffy~\cite{kobus2022jiffy}.}
Jiffy is a multi-versioned lock-free skiplist. Similar to JellyFish~\cite{yeon2020jellyfish}, Jiffy  keeps a vertical list of revisions at each entry in the bottom layer. Each revision has a version number, and contains multiple key-value pairs in the range from the rooted skiplist entry and the next skiplist entry. Revision uses  copy-on-write, and is installed atomically to the list via compare-and-swap.  Obsolete revisions are garbage collected. Jiffy  supports batch updates. Updates are batched and are performed atomically. All update operations are stored in one batch descriptor. 
The revisions created by the batch updates reference the batch descriptor. A node covering a smaller range has smaller revisions that are better for write-heavy workloads while the larger ones are more suited for read-heavy workloads. Adaptation to workloads is performed by adjusting the node size. Jiffy uses  autoscaling  for  dynamic adjustments. By monitoring the ratio of time spent on reading and updating, Jiffy decides an optimal size of a node and performs node merges (so that a node covers larger ranges) or splits (so that  nodes cover smaller ranges) accordingly. We term this skiplist  \texttt{Jiffy} (Section~\ref{subsubsec:cc-perf-cmp}).

\noindent
{\bf The Frugal Skiplist~\cite{kim2021frugal}.}
The frugal skiplist  speeds up version lookup that is part of vWeaver~\cite{kim2021frugal}, an access method that allows  fast scan over records. The frugal skiplist organizes the versions that belong to the same key and the per-key search structure is weaved for an efficient move to the nearby key's neighboring version~\cite{kim2021frugal}. There are two types of pointers: \textit{next} points to the predecessor node in the same vertical stack, and the \textit{v\_ridgy} pointer is a shortcut link connecting to the node that is higher in level. To insert a new version as a node, it is added in the ground level or stacking on top of the current level as in Figure~\ref{fig:frugal} left. \textit{v\_ridgy} is formed between the new node and the node on top of the previous stack whose level is no smaller than the inserted node (Figure~\ref{fig:frugal} right). This implies an exponential growth in the number of skipped nodes.

\subsubsection{Performance Comparison}\label{subsubsec:cc-perf-cmp}
Next, we show the performance comparison between the skiplists and the tree counterparts as well as the effect of different techniques applied to the above skiplists.

\noindent
{\bf 1. Comparison of Skiplists.}
In~\cite{pugh1998concurrent}, there is no comparison between \texttt{BasicSL} and its tree counterparts as the skiplist is not well suited for disk-based data structures~\cite{pugh1998concurrent}.  \texttt{BasicSL} scalability is tested by a simulated experiment. Up to 1000 threads are simulated, each inserting and deleting a random element. No two threads attempt to insert or delete the same element simultaneously. For 1000 threads, the simulated speed-up is 921, almost scaling linearly. In~\cite{herlihy2007simple, herlihy2006provably}, the skiplist properties are preserved, maintaining a correct skiplist. The experiments compare \texttt{LazySL} with  \texttt{JavaImpl} and a sequential skiplist (baseline). Under two lookup-heavy workloads (9\% insert, 1\% delete, 90\% lookup; 20\% insert, 10\% delete, 70\% lookup), \texttt{LazySL} is slightly better 
for a smaller-range dataset. For a high contention workload (50\% insert, 50\% delete), \texttt{LazySL} degrades  as the failed validation does not have a back-off mechanism. In~\cite{platz2019concurrent}, the correctness of the unrolled skiplist algorithm is proven.  The experiments compare \texttt{UnrolledSL} with \texttt{LazySL}, \texttt{FraserSL} and \texttt{CFSL} for uniform and Zipfian datasets. Three workload are used: read-biased: 90\% lookup, 5\% insert and 5\% delete; balanced: 50\% lookup, 25\% insert, 25\% delete; and write-biased: 0\% lookup, 50\% insert and 50\% delete. \texttt{UnrolledSL} with  group mutual exclusion  performs the best, followed by \texttt{CFSL}, \texttt{FraserSL} and \texttt{LazySL}. In~\cite{fraser2004practical}, Fraser's designs are evaluated.
The STM-based implementation is outperformed by the other lock-free schemes. For low contention, the MCAS-based design has almost identical performance to the CAS-based design. In~\cite{crain2013no}, the No Hotspot contention-friendly skiplist (\texttt{CFSL}) is compared with \texttt{JavaImpl} using a main-memory database benchmark. \texttt{CFSL} has up to 2.5$\times$ increase in throughput when the update ratio is high. \texttt{CFSL} scales better with the increase in the number of threads. In~\cite{dick2017rotating}, \texttt{RotateSL} is compared against \texttt{FraserSL} and \texttt{CFSL}. \texttt{RotateSL} scales better with more threads and is more tolerant to contention that is caused by a high update ratio. In~\cite{yeon2020jellyfish}, \texttt{JellyFish} is implemented  in RocksDB~\cite{rocksdb} as JFDB and is compared with the original RocksDB.  JFDB outperforms RocksDB by 8\%, 16\%, 51\%, and 545\% for insert, update, lookup, and range query, respectively, under high concurrency (16 threads).

\noindent
{\bf 2. Comparison With Other Data Structures.}
In~\cite{kobus2022jiffy}, \texttt{Jiffy} is compared with several contention-adaptive (CA) data structures CA-imm~\cite{sagonas2015contention}, CA-AVL~\cite{sagonas2018contention}, CA-SL~\cite{sagonas2018contention}, LFCA tree~\cite{winblad2021lock} and SnapTree~\cite{bronson2010practical}, k-ary tree~\cite{brown2011non,brown2012range}.A CA data structure can adapt the structure based on the collected contention statistics~\cite{sagonas2015contention,sagonas2018contention}. \texttt{Jiffy} has a lower throughput in single put/remove operations than SnapTree, CA-imm, LFCA tree or CA-AVL, but a higher range scan throughput than the compared indexes. \texttt{Jiffy} performs better than the compared indexes for larger batch update sizes.

\subsection{Enforcing Determinism}\label{subsec:deterministic}

In this section, we discuss the search, insert, and delete operations in 1-2 deterministic skiplists~\cite{munro1992deterministic}.
Network overlay and maintenance algorithms, e.g.,~\cite{clouser2008tiara, nor2013corona, harvey2002skipnet, harvey2003brief, singh2015concurrent, mandal2012deterministic}, widely use  deterministic skiplists. 
Dynamic overlay networks are applied in peer-to-peer systems. As the skiplist supports logarithmic search and update, and can also be made deterministic, the skiplist has become a valid candidate for network overlay algorithms~\cite{clouser2008tiara,singh2015concurrent,mandal2012deterministic, nor2013corona} (More discussion   in Section~\ref{sec:more-use}).

\noindent
{\bf 1-2 The Deterministic Skiplist~\cite{munro1992deterministic}.}
Search in the 1-2 deterministic skiplist is the same as that in the probabilistic skiplist. One advantage in the 1-2 skiplist is that, during search, there is  descendance of one level after every two horizontal steps~\cite{munro1992deterministic}. The key point is to ensure the gap invariant that between any two consecutive nodes at Height $h$, there are at most 2 nodes at Height $h-1$~\cite{munro1992deterministic}.
An insert in the 1-2 skiplist first adds the item at Height 1 (bottom level). This may invalidate the gap invariant by having a third item of Height 1 in a row. Thus, the middle item is promoted to Height 2. This repeats until reaching the skiplist's maximum height~\cite{munro1992deterministic}. To maintain the gap invariant during deletes, demoting or promoting a node may be needed~\cite{munro1992deterministic}, e.g., in Figure~\ref{fig:1-2-3_skiplist}, deleting Node 5 promotes Node 3 to a higher level, and the gap invariant is still intact.

In contrast to a probabilistic skiplist, 
the height of a node in a deterministic skiplist is not decided by consulting a random number generator, rather by maintaining the gap invariant.
In the 1-2 deterministic skiplist of $n$ different items, the number of horizontal pointers in the worst can never exceed $2n$~\cite{munro1992deterministic}. Since the insert process involves items to grow, this can cause copying of horizontal pointers if they are stored in an array. Then, an insert can take up to $\Theta(\log^2n)$ time~\cite{munro1992deterministic}. If horizontal pointers are stored in a linked list, this reduces the time complexity but increases  storage by up to $6n$. Munro et al.~\cite{munro1992deterministic} use an array of exponentially increasing heights and the allocated space is smaller than $2.282n$ pointers. The time complexity of inserts is $\Theta(\log n)$.

Both the probabilistic and the deterministic skiplists use active inserts, and deletes, meaning that updates are immediately propagated to the index layers, which makes these structures  balanced.

\subsection{Accommodating for Skewed Data Access Patterns in Skiplists}\label{subsec:data-access}

Access patterns are often skewed, e.g., locality of reference in memory~\cite{tanenbaum1997operating}, disk~\cite{o1993lru} and buffer management~\cite{dan1990approximate}. Many data structures have been proposed to deal with these access patterns. 

\noindent
{\bf The Biased Skiplist (BSL)~\cite{ergun2001biased}.}
Ergun et al.~\cite{ergun2001biased} introduce a skiplist for biased access patterns. The skiplist is constructed as an ordinary skiplist except that the keys are partitioned into {\em classes}, and data is copied either automatically or randomly into the upper levels. 
The motivation for replicating some keys is that the frequently accessed keys can be replicated and stored in the higher levels in the skiplist and hence are found faster. 
To assign keys to  classes, keys are ordered in ascending order by their rank $r(k)$; the number of distinct keys accessed since the last access to $k$. This ordered list is partitioned into classes $C_1$, $C_2$, ..., $C_{\log n}$ contiguously where class size is $|C_i| = 2^{i-1}$. Class 1 is the smallest rank, which suggests that the data should be accessed early in the skiplist's upper levels. Upon construction, the key height is determined by its class and by a random number (details in~\cite{ergun2001biased}). Searching a key is the same as in an ordinary skiplist. After the key is found, the rank of the key becomes 1, and it is moved to the front of the rank ordered linked list. Keys in other classes are moved to different classes due to the class size constraint. In BSL, the expected search time for key $k$ is $O(\log r(k))$. In a sense, this mimics Shannon's coding theory~\cite{shannon1948mathematical}, where frequently accessed items get assigned a shorter code length. Similarly, in BSL, a frequently accessed item, say $i$, will require less probes to the skiplist before $i$ is found, the reason being that $i$ will be replicated into the upper levels in the skiplist as $i$ gets accessed more frequently, and hence is found faster.
In~\cite{ergun2001biased}, BSL is compared against the basic skiplist and a binary trie. BSL outperforms the basic skiplist and the trie on 64-bit keys on the tested bias (i.e., the average rank of the searched keys), 
but cannot outperform the trie on 32- and 48-bit keys under some bias.

\noindent
{\bf The ($a, b$)-Biased Skiplist And Randomized Biased Skiplist~\cite{bagchi2005biased}}. 
Bagchi, Buchsbaum, and Goodrich~\cite{bagchi2005biased} introduce 
 a new biased skiplist, where each key is assigned a weight $w_i$. The goal is to achieve faster search time than $O(\log n)$ for highly weighted items. Bagchi et al. devise two biased skiplists: deterministic and  randomized. The ($a, b$)-biased skiplist is  deterministic, where the height of the item $h_i \geq r_i$ and $r_i$ is the rank defined as $\lfloor \log_aw_i \rfloor$. Two invariants are enforced: (1)~There are no more than $b$ consecutive items at Height $i$; and (2)~For Node $x$ and all $i$ that are between $x$'s rank and height, there are at least $a$ nodes of Height $i-1$ between $x$ and any consecutive node of Height at least $i$. The ($a, b$)-biased skiplist accesses an item in $O(1+\log (W/w_i))$ time in the worst case~\cite{bagchi2005biased}. The randomized version assigns the height of an item to be the sum of the rank-determined number and a random number. It can achieve the above bound in the expected case~\cite{bagchi2005biased}.

\begin{figure*}[h]
    \centering
    \includegraphics[width=0.78\linewidth]{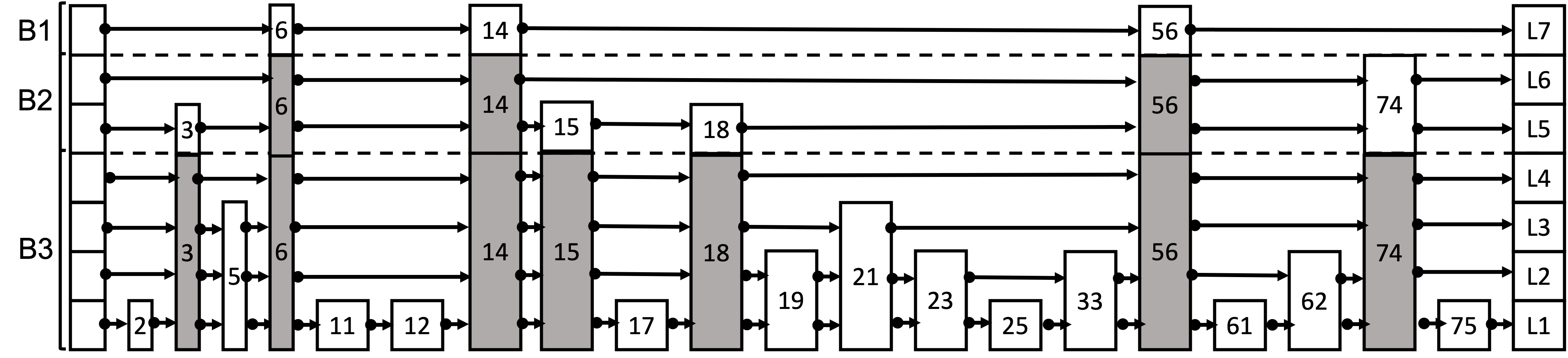}
    \caption{Self-adjusting skiplist (SASL)}\label{fig:sasl}
\end{figure*}

\noindent
{\bf The Self-Adjusting Skiplist (SASL)~\cite{ciriani2002static, ciriani2007data}}. SASL is a skiplist that can promote or demote its node's height. The height of a node, say $s$, in an $n$ data-item skiplist is bounded by $H=\Theta(\log n)$. SASL  uses the access pattern 
to determine the height of a node, similar to~\cite{ergun2001biased}. The horizontal linked lists of SASL are grouped into multiple bands similar to the classes in~\cite{ergun2001biased}. The size of the band grows exponentially. 
Refer to Figure~\ref{fig:sasl}. Assume  there are 20 data items in the skiplist and 3 bands. $B_1$ has List $L_7$ while $B_2$ has Lists $L_5$ and $L_6$, etc. Each item is associated 2 two integers, a random value $r$ and a deterministic value $d$. $r$ is the random height within a band and $d$ is the height of some band $B_i$. A node, say $s$, resides in $B_i$ if $d(s)=H(i)$. In Figure~\ref{fig:sasl}, if $r(15)=1$ and $d(15)=4$, Key 15 is in $B_2$ as $H(2)=4$. The shadowed part of Key 15 in Figure~\ref{fig:sasl} is the deterministic part that is Height 4. The height of Key 15 inside $B_2$ is $r(15)=1$. 
Search proceeds as in the ordinary skiplist. Adjustment is performed afterwards. If the target element $x$ is in $B_i$, $x$ is promoted to $B_1$. $r(x)$ is not discarded for future demotion. For each $B_j$,  $j \in [1, i-1]$, a random element is chosen to be demoted from $B_j$ to $B_{j+1}$. This search and restructuring has $O(2^i)$ expected time~\cite{ciriani2002static, ciriani2007data}.

\noindent
{\bf The T-List~\cite{mei2017concurrent}.}
The T-list skiplist  is constructed and modified during search. Its intuition is similar to the 1-2 deterministic skiplist~\cite{munro1992deterministic} where  search can only visit  a fixed number of nodes in the same level.  Mei et al.~\cite{mei2017concurrent} define 
a fixed \textit{span} to limit the number of horizontal steps. Unlike the deterministic skiplist, the T-List does not enforce a strong constraint. 
During search, when the number of consecutive horizontal steps exceeds a certain threshold \textit{span}, that node is promoted to a higher level. 
For example, let \textit{span} be 2. If searching a key requires 3 horizontal steps in Level-0 that exceeds the threshold 2, the third node is promoted to a higher level, and thus can be located in fewer steps. 
To promote a node,  two nodes need to be locked: the current modifying node and the first node in this level before the current node. This can increase concurrency. Logical delete is performed by inserting an invalidated entry.
The T-list is compared with the skiplist~\cite{mei2017concurrent}. 
With \textit{span=2}, the T-list is better for all workloads. When \textit{span=4}, the T-list is worse for list sizes $10^5$ and $10^6$.

\begin{figure*}[h]
\centering
    \hspace*{\fill}%
    \begin{subfigure}{0.3\textwidth}
    \centering
        \includegraphics[width=\linewidth]{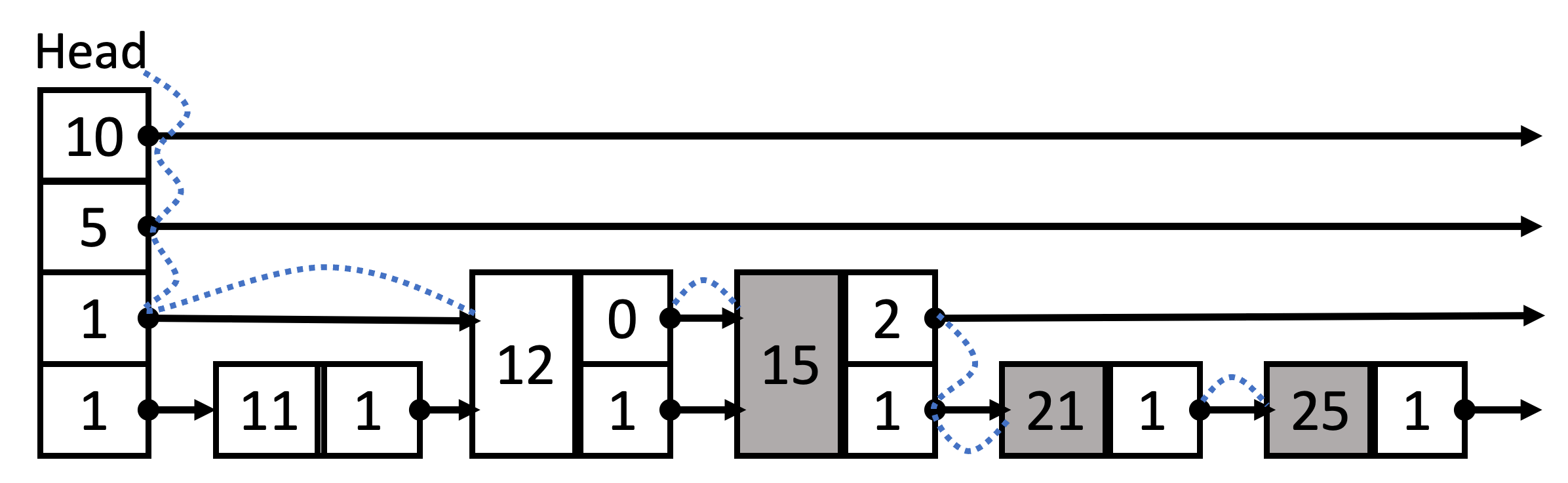}
        \caption{The splay-list}\label{fig:splaylist}
    \end{subfigure}
    \hfill
    \begin{subfigure}{0.2\textwidth}
    \centering
        \includegraphics[width=\linewidth]{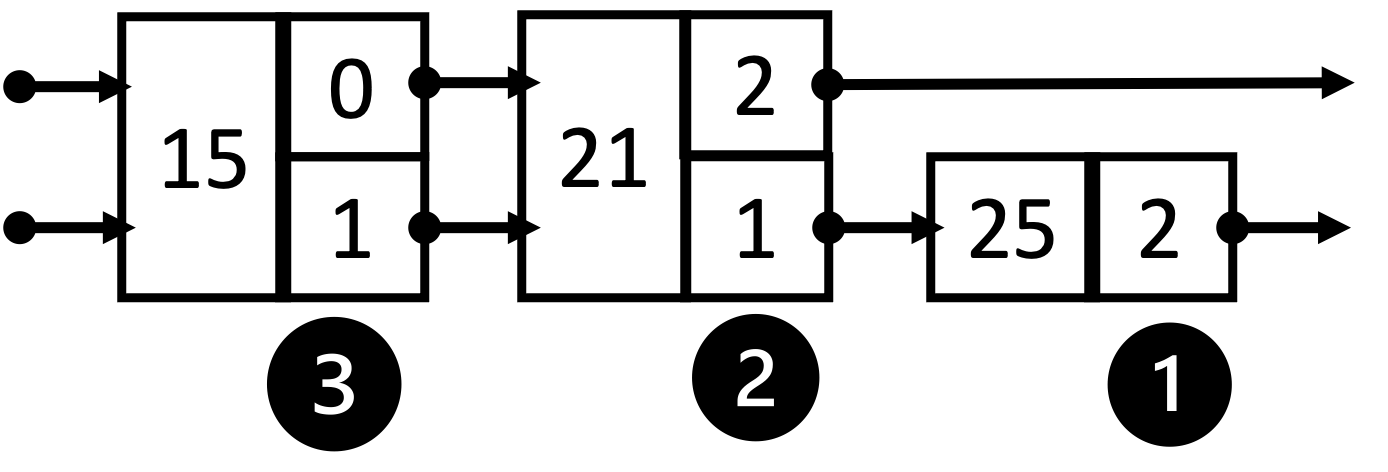}
        \caption{Steps 1-3}\label{fig:splaylist-step-1}
    \end{subfigure}
    \hfill
    \begin{subfigure}{0.2\textwidth}
    \centering
        \includegraphics[width=\linewidth]{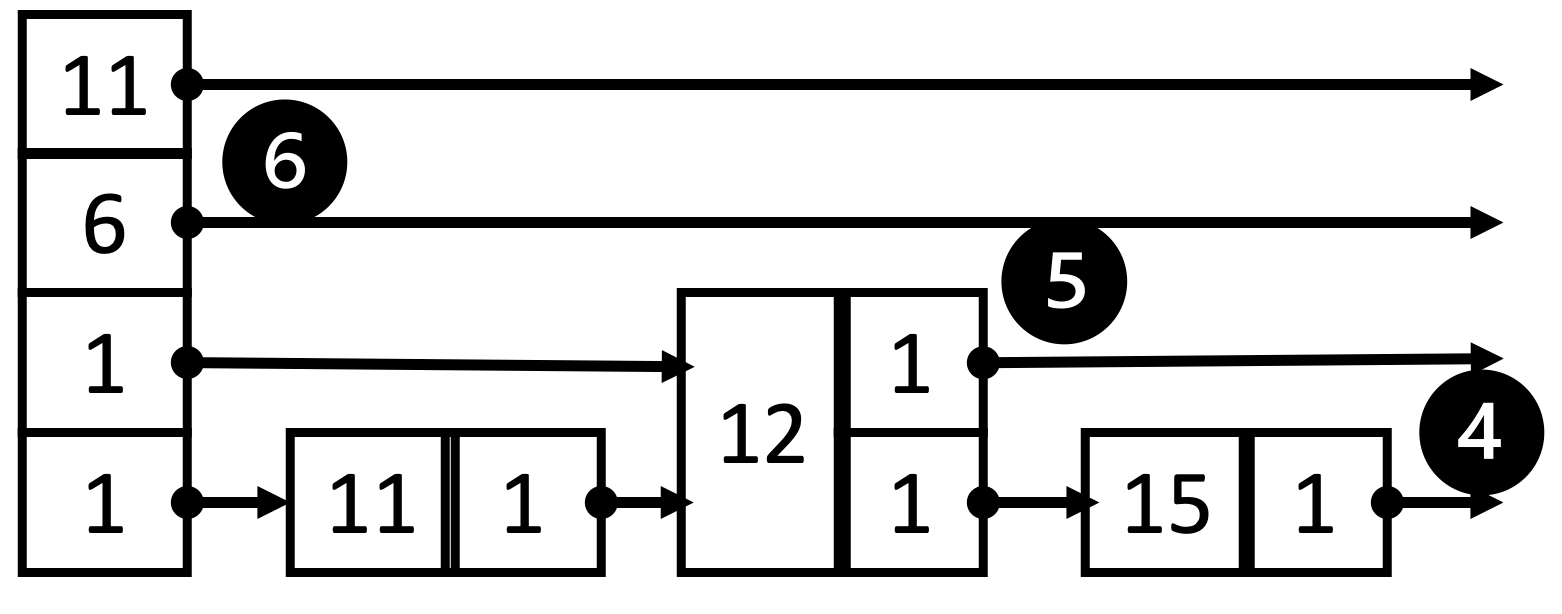}
        \caption{Steps 4-6}\label{fig:splaylist-step-2}
    \end{subfigure}
    \hspace*{\fill}%
\caption{The splay-list (data $>$ 25 is not shown). Promotion/demotion after the operation \textit{contains(25)}.}\label{fig:skiplist-tree}
\end{figure*}

\noindent
\noindent{\bf The Splay-list~\cite{aksenov2023splay}.}
The Splay-list is a skiplist constructed based on data access frequency. The  frequently accessed nodes are promote to higher levels in the skiplist, and the less frequently accessed nodes are demoted. Let $h_u$ and  be the maximum height of a node $u$, and let $C_u^h$ be the children nodes of $u$ at Height $h$, which is also $u$'s subtree at $h$. The subtree of $u$ includes the successor nodes of $u$ at Height $h < h_u$, e.g., the subtree for Node 15 contains Nodes 15, 21 and  25 (shaded nodes in Figure~\ref{fig:splaylist}). 
In Figure~\ref{fig:splaylist}, both the head and tail store the number of accesses of the subtree (tail is not shown). Let $H$ be the head. For each node, the left box stores the key, and the right box stores the number of accesses to the subtree.
The height starts at 0. In the example, $C_{15}^1 = \{15, 21, 25\}$; $C_{12}^1 = \{12\}$, $C_{H}^1=\{H, 11\}$, $C_{H}^2=\{H, 11, 12, 15, 21, 25\}$ including the subtree of Node 15. Assume that the total number of accesses to the skiplist is $m=10$. Thus, the height of the tree is $\lfloor \log (m)\rfloor=3$. For the descent condition, we compute the sum ($C_1$) of accesses of the subtrees of the neighboring nodes $u$ and $v$. If the sum is less than a descent threshold, the node is demoted,  and is merged into the subtree of its left predecessor. Similarly, for the ascent condition, we compute the sum ($C_2$) of accesses of the subtree. If it is greater than a threshold, the node is promoted. In the example, after \textit{contains(25)}, Key 25 is found, and the skiplist is adjusted in a backward pass. First, we increment the access number for Node 25. To decide whether to demote Node 21, $C_1=C_{15}^0+C_{21}^0=2$, which is below the threshold. For Node 21, $C_2 = C_{21}^0+C_{25}^0=3$  exceeds the threshold, and hence  is promoted to Level-1. Thus, the skiplist becomes as in Figure~\ref{fig:splaylist-step-1}. The access number of Node 15 at Level-1 is set to 1. Similar adjustments are performed to Node 15 as in Figure~\ref{fig:splaylist-step-2}. 
In~\cite{aksenov2023splay}, the splay-list is compared with the basic skiplist and CBTree~\cite{afek2014cbtree}, a concurrent data structure leveraging data skew. In the single-threaded experiment, the splay-list outperforms the basic skiplist under fewer updates. The splay-list outperforms when the skew is high and rebalancing is performed more frequently.


\subsection{Skiplist Partitioning}\label{subsec:partition}
Partitioning the skiplist into multiple disjoint ranges is more straightforward than its tree counterparts. This optimization has been widely adopted in indexing in distributed systems, network overlay algorithm etc. Furthermore, partitioning the skiplist into smaller ranges can lower the skiplist height that reduces the number of random accesses and the number of pointer updates.

\subsection{Changing Node Content}\label{subsec:node-content}
In the basic skiplist, a node contains one data item and a variable number of pointers. We present two optimizations: Limiting the number of pointers, and storing multiple data items per node.

\subsubsection{A Fixed Node Size Skiplist}
\hfill

\noindent
{\bf The Modified Skiplist. }
Cho and Sahni propose a modified skiplist~\cite{cho1998modified}. In the basic skiplist, a node can have a variable number of pointers. Refer to Figure \ref{fig:skiplist_search}. Node 11 has 1 pointer while Node 21 has 3. The modified skiplist has $O(1)$ pointers where each node contains 3 pointers: \texttt{left, right, down}. The \texttt{left} and \texttt{right} pointers make each level a doubly-linked list. The \texttt{down} pointer of Node $A$ on Level-$h$ points to the node that is one level down (Level-$(h-1)$) that is the smallest value that is larger than $A$. The complexity of the modified skiplist is at most logarithmic extra work compared to the basic skiplist~\cite{cho1998modified}. Emperically, the modified skiplist uses lesser time than the basic skiplist~\cite{cho1998modified}. The modified skiplist requires less storage if implemented in programming languages that do not support dynamic construction of variable-size arrays, but requires more storage if implemented in \texttt{C/C++}, \texttt{Java} that support  dynamic construction of variable size arrays~\cite{cho1998modified}.

\subsubsection{Unrolling Node}\label{subsubsec:unrolled}\hfill

\noindent
Visiting a skiplist node requires multiple pointer traversals. Pointer chasing may not be optimal for caches and on-disk structures. Several skiplist variants group multiple data items into one node. This grouping is termed {\em unrolling}. It reduces the number of pointers in the skiplist, and thus improves cache performance. Since the number of nodes is reduced, the number of promotions/demotions during skiplist update is reduced. 
This technique has been applied to improve cache locality in cache-sensitive skiplists~\cite{sprenger2017cache}, write performance in non-volatile storage-based  skiplists\cite{bender2017write, wang2017flashskiplist, li2022phast}, and range search performance\cite{avni2013leaplist}. Concurrency in the unrolled skiplist~\cite{platz2019concurrent} is studied in Section~\ref{subsubsec:lock-based}.


\section{Variants of the Skiplist: Taxonomy and Related Systems}
\label{sec:taxonomy}

\begin{figure}[h]
    \centering
    \includegraphics[width=\linewidth]{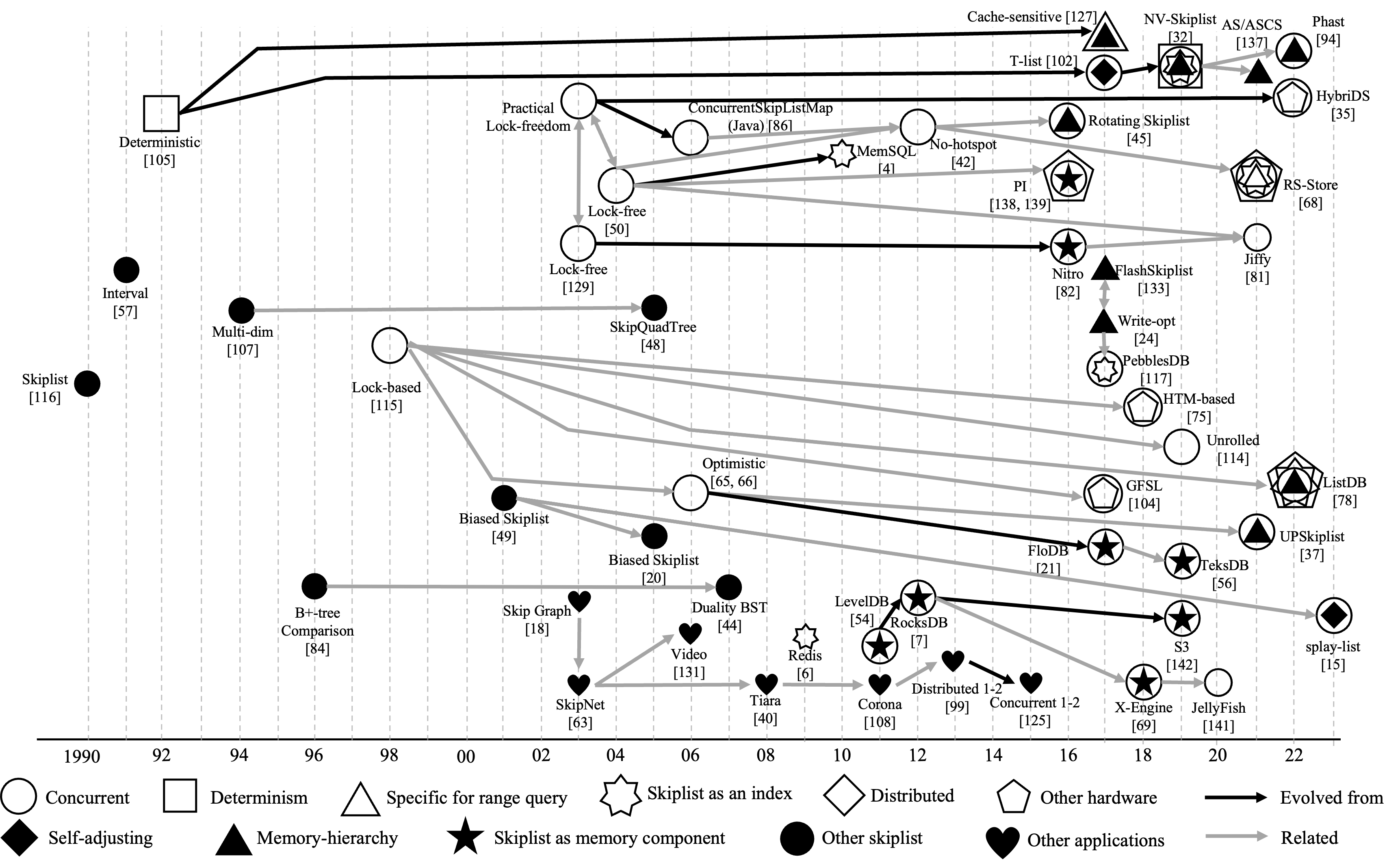}
    \caption{The taxonomy of skiplist variants and their applications.}
    \label{fig:taxonomy}
\end{figure}

Figure~\ref{fig:taxonomy} illustrates the evolution of the skiplist variants over time, and the various systems that make use of these variants. All skiplist variants originate from the first skiplist~\cite{pugh1990skip}. Thus, we omit the connection from the first skiplist paper. 
\vspace{-\topsep} 
\begin{enumerate}
    \item {\bf Skiplists as Data Indexes or Storages:} The equivalence between the skiplist and the B/B$^+$-tree is highlighted in Section~\ref{sec:contrast-btree}. Thus, skiplists are used often in database systems. 
    The initial skiplist~\cite{pugh1990skip} is designed as an in-memory index. Later, the skiplist is adapted to fit  modern hardware.  We study these skiplist variants  in Section \ref{sec:hardware}. Furthermore, as skiplists can be used in various hardware platforms, this enables a broader use of skiplists in various big data systems to either store and index the underlying data. This is studied in Section~\ref{sec:kvstore}.
    
    \item {\bf Supportive Role of Skiplists:} 
    Many key-value stores use Log-Structured Merge Tree (LSM-Tree) to store data. In this context, the skiplist is frequently used as the LSM-Tree's memory component. We  describe the integration of skiplists in LSM-Trees in Section~\ref{subsec:integrating}. Also, skiplists can implement priority queues. These skiplist variants are presented in Section~\ref{subsec:priority-queue}.
    \item {\bf Indexing Complex Data: } 
    Skiplists have been used to store or index other data types, including multi-dimensional data and datq intervals (Section~\ref{sec:complex-data}).
    \item {\bf Skiplists In Operating Systems:} The skiplist has been investigated for use in the Linux kernel since 2001. Its is used in place of red-black trees. One skiplist implementation outperforms the red-black tree in a multi-threaded testing~\cite{skiplistvsrbtree}. MuQSS, a process scheduler, is also based on the skiplist~\cite{muqss}. A more detailed kernel skiplist investigation can be found in~\cite{kernelskiplist}.
    \item {\bf Efficient Range Queries with Skiplists:} 
    Skiplists maintain sorted data. This   allows fast retrieval of ranges of data. We present skiplists supporting efficient range queries in Section~\ref{sec:range-query}.
    \item {\bf The Extensibility of Skiplists: }
    The heuristics used for the skiplist can be applied to other data structures and algorithms. We describe these in Section~\ref{sec:extensibility}.
    \item {\bf More Uses of the Skiplist: }
    Section~\ref{sec:more-use} describes how skiplists are used in other domains.
\end{enumerate}


\section{Optimizing Skiplists for Modern Hardware}\label{sec:hardware}
In this section, we explore the implications of new hardware on the designs of the skiplist. This includes the impact of new types of memory including flash and persistent non-volatile memories, and CPU architectures including NUMA awareness and GPUs.

\subsection{Improving Cache Locality}\label{subsec:cache}
\noindent
{\bf The Cache-sensitive Skiplist~\cite{sprenger2017cache}.}
The cache-sensitive skiplist~\cite{sprenger2017cache} has been introduced as the linked list offers poor spatial locality. The nodes of a linked list are not necessarily located contiguously next to each other in memory. This results in a cache miss every time a pointer is chased.
In order to address this issue, Sprenger, Zeuch, and Leser~\cite{sprenger2017cache} use a linearized fast lane that stitches the entire index layer into a single array. The cache-sensitive Skiplist eliminates the use of pointers, and calculates the positions of the child nodes by using arithmetic. This saves space, and reduces cache misses. Moreover, SIMD instructions~\cite{10.1145/564691.564709} can be used to further speed up operations. These  benefits have the disadvantage that the index layers become rigid, i.e., whenever an insert occurs, the node is only inserted in the data layer, and after the number of inserts reaches a threshold, the cache-sensitive skiplist reconstructs the new index layers. Similarly, for a delete operation, the node is deleted in the data layer, 
and all the corresponding positions in the index layers with the smallest value greater than the deleted node are updated. This is performed to avoid false positives during search in the cache-sensitive skiplist. The cache-sensitive skiplist cannot use null value, as it cannot be compared against the search value during traversals, so the only logically valid option is to copy the next greater value. 
In~\cite{sprenger2017cache}, the cache-sensitive skiplist (CSSL) is compared with the adaptive radix tree (ART)~\cite{leis2013adaptive}, the cache-sensitive B$^+$-tree~\cite{rao2000csb} and a B$^+$-tree~\cite{bayer1970organization}. For range query workloads with 16M and 256M  32-bit integer keys, CSSL has superior  performance. Under a lookup workload, ART performs the best for 16 million 32-bit keys. Under a mixed workload of 50\% lookup queries and 50\% range queries, CSSL has superior performance.

\subsection{External Memory Storage}
\label{subsec:external-mem}

\subsubsection{Flash Memory}\label{subsubsec:flash}
Flash Memory has asymetric read and write latencies where reads are faster than writes. Also, flash memory has the erase-before-write constraint that increases the write cost. 

\begin{figure*}[h]
\centering
\hspace*{\fill}%
    \begin{subfigure}{0.3\textwidth}
    \centering
        \includegraphics[width=\linewidth]{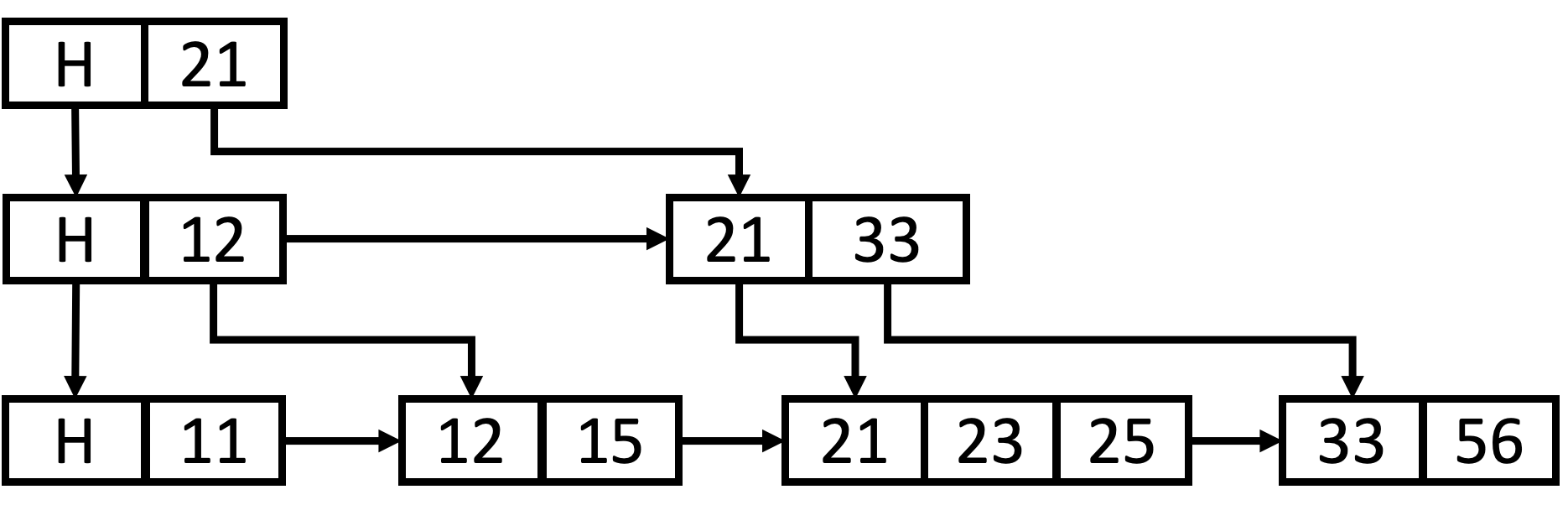}
        \caption{Tree representation of  skiplist}\label{fig:tree-representation}
    \end{subfigure}
    \hfill
    \begin{subfigure}{0.4\textwidth}
    \centering
        \includegraphics[width=\linewidth]{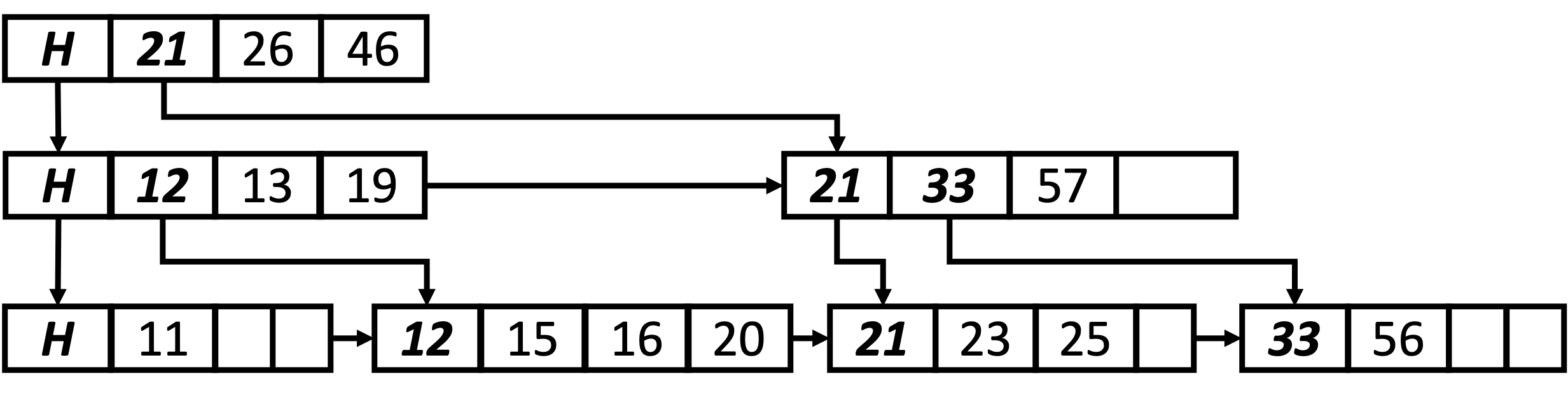}
        \caption{The write-optimized skiplist}\label{fig:wo-skiplist}
    \end{subfigure}
\hspace*{\fill}%
\caption{The write-optimized skiplist.}\label{fig:skiplist-tree}
\end{figure*}

\noindent
{\bf The Write-optimized Skiplist~\cite{bender2017write}.}
Bender et al.~\cite{bender2017write} present a write-optimized skiplist that works on external-memory of block size $B$ to achieve asymptotically better insertion performance than the B-tree and offer similar range search performance. Each node contains pivots and a buffer, but the amount of space assigned to each is variable. Based on the promotion probability~\cite{bender2017write}, on Level $L_{i\geq 1}$, each node has $\Theta(B^{\epsilon})$ pivots in expectation, and on Level $L_0$, there are expected $\Theta(B^{1-\epsilon})$ pivots. The smallest pivot per node is termed \textit{leader}. In Figure~\ref{fig:wo-skiplist}, the bold elements are pivots and the number of pivots per node is different. All the data items in the buffer are greater than the leader of the same node, but are smaller than the leader of the successor node in the same level. Figure~\ref{fig:tree-representation} gives the same skiplist as the one in Figure~\ref{fig:1-2-3_skiplist} but in a different representation. Insertions are first buffered in the buffer area in the node until the buffer overflows. A flush operation is triggered to distribute each data item in the buffer to the corresponding child node. If the height, say $h_e$, of data item $e$ is greater than the current level, this suggests that $e$ needs to become a pivot in the lower levels. This involves node splitting and establishing new parent-child node link. Insertions take $O((\log_{B^{\epsilon}}N/(B^{1-\epsilon})$ amortized I/Os in expectation and with high probability. Its range search returning $K$ elements achieves $O(\log_{B^{\epsilon}}N + K/B)$ I/Os w.h.p for $0 < \epsilon < 1$ and block size $B$~\cite{bender2017write}.

\begin{figure*}[h]
\centering
    \begin{subfigure}{0.49\textwidth}
    \centering
        \includegraphics[width=\linewidth]{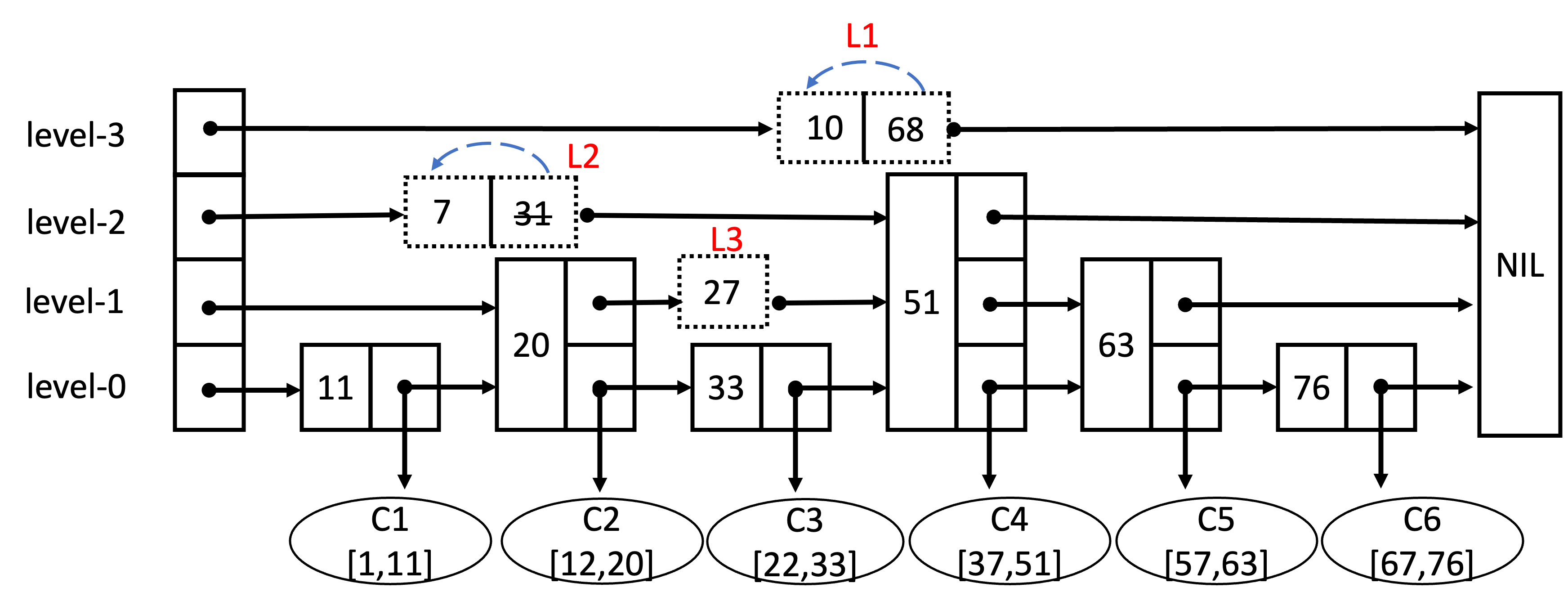}
        \caption{The FlashSkipList}\label{fig:flash_skiplist}
    \end{subfigure}
    \hfill
    \begin{subfigure}{0.49\textwidth}
    \centering
        \includegraphics[width=\linewidth]{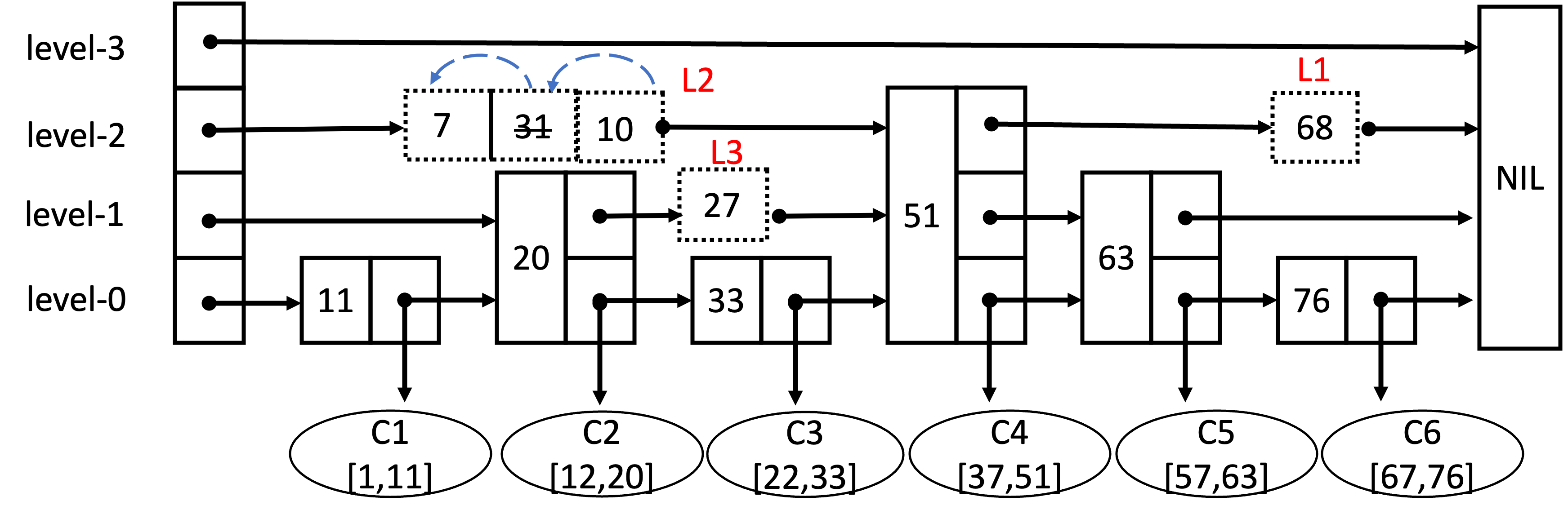}
        \caption{FlashSkipList Rearragement of List L1. 
}\label{fig:flash_skiplist_rearrangement}
    \end{subfigure}
\caption{The FlashSkipList and its operation}\label{fig:flash}
\end{figure*}

\noindent
{\bf The FlashSkiplist~\cite{wang2017flashskiplist}.}
The FlashSkiplist uses an unrolled node structure (Section~\ref{subsubsec:unrolled}).
Entries are grouped in chunks that can be accessed through skiplist nodes. The skiplist is constructed based on the largest entry per chunk. This design saves the space for pointers, and is I/O efficient as one I/O brings in multiple entries. Also, batched updates can be applied to the same chunk at once~\cite{wang2017flashskiplist}.
The FlashSkiplist data structure~\cite{wang2017flashskiplist} tackles the constraints posed in flash memory by using a write-optimized component, a read-optimized component, and a dynamic rearrangement strategy.
The FlashSkiplist buffers the writes into the write-optimized component that uses an append-only strategy to achieve high write throughput. As  in Figure~\ref{fig:flash_skiplist}, lists $L1-L3$ comprise the write optimized component. Any new updates will be appended to the lists in the topmost level of the FlashSkipList.
The read-optimized component is comprised of chunks of pages. As  in Figure~\ref{fig:flash_skiplist}, the chunks $C1-C6$ form the read-optimized component. Each node in the flash skiplist points to a chunk, and the value of the node corresponds to the maximum value allowed in the particular chunk.
The FlashSkiplist uses the dynamic rearrangement that restricts the length of the write-optimized list so that the read cost is manageable. The writes can be batched and transferred to the read-optimized skiplist
component to decrease the overall write costs. 
An insert operation is simply an append to lists in the topmost level of the FlashSkiplist. As  in Figure~\ref{fig:flash_skiplist}, when a new item 68 arrives, it is appended to the list $L1$ in the topmost level of the FlashSkiplist.
A delete operation inserts a ghost entry into the top level list. The ghost entry is an entry that indicates that the element has been deleted. For example, as in Figure~\ref{fig:flash_skiplist}, in List $L2$, Element \sout{31} is a ghost element, meaning that 31 has been deleted. 
A search operation takes places in two steps. The first step is a skiplist search that finds the corresponding list or chunk in which the key is present. The second step is to search within a linked list or a chunk.
Rearrangement is performed on a single list. Rearrangement of a list $L_i$ will cause each element of the list to be pushed to a lower level. As  in Figure~\ref{fig:flash_skiplist_rearrangement}, rearranging the list $L1$ causes Element 10 to be appended to List $L2$ and Element 68 to be pushed to Level 2. The rearrangement of a list in Level 1 causes each element of the list to be written to the chunk. For example, if List $L3$ is pushed, element 27 will be written to Chunk $C3$.
The FlashSkipList is evaluated against three indexes designed for flash memory: A B-tree index BFTL~\cite{wu2007bftl}, a self-tuning index in FlashDB~\cite{nath2007flashdb}, a lazy-adaptive LA-tree~\cite{agrawal2009latree}, and a tree structure FD-tree~\cite{li2010fdtree}. The FlashSkipList shows the smallest response time under uniform workload ($1.1 - 5.5\times$), a real-world workload \textit{Weather} ($1.2 - 14.3\times$) and TPC-C workload ($1.1 - 5.1\times$).


\subsubsection{Persistent Memory}\label{subsubsec:PM}
Intel\textregistered Optane\textsuperscript{TM} DC Persistent Memory (PM, for short), also termed Non-Volatile Memory (NVM), is a class of storage that is both byte-addressable and persistent~\cite{izraelevitz2019basic}. To guarantee data durability in PM, persistence barrier including cacheline flush plus store fence, is needed. In order to fit the skiplist into PM, the traditional skiplist needs to be modified such that any updates to the skiplist are persisted immediately to the PM storage~\cite{yang2020empirical}. Oukid et al.~\cite{oukid2015instant} compare a persistent skiplist with a traditional DRAM-based skiplist. They find that both read and write performances degrade in the persistent skiplist because of the high latency of PM and the random access pattern of skiplist. Efforts have been put into designing performant persistent skiplists in recent years~\cite{chen2019design, lee2019recipe, chowdhury2021scalable, xiao2021write, li2022phast}. 
We give an overview of persistent skiplists below.

\noindent
{\bf  The NV-skiplist~\cite{chen2019design}.}
The NV-skiplist~\cite{chen2019design} is a persistent skiplist that has been tested on a PM emulator.
The NV-skiplist adopts designs that reduce the PM writes: (1)~Multiple entries are grouped into one node, and remain unsorted within the node. (2)~A bitmap is used to identify empty slots. (3)~Selective persistence 
where only the last level is persisted
in PM while the internal levels are stored in DRAM to keep only last level consistent and rebuild the internal levels upon system failure. The NV-skiplist has introduced two optimizations. 
The first optimization uses the deterministic design~\cite{mei2017concurrent} to optimize search performance. 
The second optimization requires the whole key space to be partitioned into several groups to increase scalability.
The NV-skiplist is compared with a B$^+$-tree designed for PM, wB+-tree~\cite{chen2015vldb}. Under uniform data, the single-threaded NV-skiplist inserts faster than the wB+-tree, and the deterministic optimization makes point read slower than the wB+-tree. With YCSB benchmark, the NV-skiplist scales better with more threads.

\noindent
{\bf The UPSkiplist~\cite{chowdhury2021scalable}.}
RECIPE~\cite{lee2019recipe} is an approach to convert DRAM-resident indexes to crash-consistent indexes for PM. These indexes follow three conditions: (1)~Write operations are visible to other threads with a single hardware-atomic store. (2)~Reads and writes are non-blocking; writes fix the inconsistencies. (3)~Reads are non-blocking, and writes are blocking as writes do not fix the inconsistencies. However, the lock-free skiplist proposed by Herlihy et al.~\cite{herlihy2020art} does not satisfy these conditions as one write operation involves multiple hardware-atomic stores, and writes are non-blocking as well as non-repairing. 
The UPSkiplist~\cite{chowdhury2021scalable} uses an epoch mechanism to solve this issue. Each node is given an \texttt{epochID}. When this epoch is different from the current epoch, the threads compete by a hardware-atomic store to fix this inconsistency. Thus, non-blocking-non-repairing operations 
in the lock-free skiplist proposed by Herlihy et al.~\cite{herlihy2020art}
are transformed into repairing ones.
The UPSkiplist is compared with the BzTree~\cite{arulraj2018bztree} that uses persistent multi-word Compare-And-Swap (\texttt{PMwCAS}) in~\cite{chowdhury2021scalable}. The UPSkiplist is outperformed by the BzTree in read-only and read-latest workload; and scales better in update-heavy and read-mostly workloads because the UPSkiplist uses a single CAS operation to update a key instead of the more complex \texttt{PMwCAS}.

\noindent
{\bf The Atomic Skiplist (AS) and  the Atomic and Selective Consistency Skiplist (ASCS)~\cite{xiao2021write}.}
Write latency is higher than read latency in PM. Moreover, excessive PM writes can result in hardware failure. To reduce PM writes, Xiao et al.~\cite{xiao2021write} use a log-free design and all PM writes are guaranteed to be atomic via the instructions \texttt{CLFLUSH} and \texttt{MFENCE}. During inserts, AS orders the pointer updates such that the skiplist is recoverable without a log: (1)~A list $L$ of skiplist pointers that need to be updated is created; (2)~A data node is created and flushed to PM; (3)~Update the pointer in this data node to point to its successor; (4)~Update the pointer of its predecessor; (5)~Update the other pointers in $L$. Since each PM write is atomic, AS is recoverable even if the system crashes during insertion. ASCS relaxes the consistency criterion in that only the bottommost list is guaranteed to be consistent while the other levels are persistent but are not necessarily consistent upon system failure. Upon recovery from a failure, internal levels are rebuilt. This results in speedy recovery compared to~\cite{chen2019design} where the internal levels are volatile.
AS and ASCS are compared against two skiplists: a Redo-Logging-based scheme for consistent and the persistent skiplist (RLS) and a non-consistency version of the persistent skiplist for the PM in~\cite{xiao2021write}. In terms of insertion and deletion latency, RLS is the slowest, followed by AS and ASCS. The non-consistent version is the fastest.

\noindent
{\bf PhaST~\cite{li2022phast}. }
The (\underline{P}artitioned \underline{H}ier\underline{a}rchical \underline{S}kipLis\underline{t}) (PhaST, for short)  is a skiplist  designed for PM. PhaST lowers the  skiplist height to reduce the random access in between layers. This is achieved by range partitioning. Each partition keeps its own skiplist index nodes. PhaST caches frequent keys in DRAM, and adopts the unrolled node structure that groups multiple data entries in one node similar to the other write-optimized skiplists~\cite{bender2017write}. But unlike traditional skiplists, PhaST keeps entries within one node unsorted. A node is augmented with a bitmap to indicate the unused slots and a fingerprint array. Writes are concurrent but node split is exclusive. To protect reads from  intermediate state, reads validate the maximum key of each node after a read  as the maximum key is the last to be updated during a split. 
PhaST is tested against the NV-skiplist~\cite{chen2019design}, the wB+-tree~\cite{chen2015vldb}, the FPTree~\cite{oukid2016fp}, FAST-FAIR~\cite{hwang2018fair} and WORT~\cite{lee2017wort}. In single-threaded experiments, under uniform data, PhaST has the highest insert throughput, has the second highest point read throughput, outperforms the NV-skiplist and the wB+-tree in scan workloads. In the YCSB benchmark, PhaST has the highest performance under workloads A, B, C and F. For multi-threading, PhaST outperforms FAST-FAIR in concurrent inserts, point read, and mixed read-write.

\noindent
{\bf ListDB~\cite{kim2022listdb}.}
ListDB presents a braided skiplist~\cite{kim2022listdb}. We overview it in  Sections~\ref{subsec:numa} and~\ref{sec:kvstore}.

\subsection{The Non-Uniform Memory Access (NUMA) Environment}\label{subsec:numa}
NUMA is the phenomenon of different memory access characteristics at different locations in the address space~\cite{lameter2013numa}. Accessing local memory  has shorter latency than accessing remote memory.

\noindent
{\bf The \underline{P}arallel In-memory Skiplist-based \underline{I}ndex (PI) and the \underline{P}arallel In-memory \underline{S}kip \underline{L}ist (PSL)~\cite{xie2016pi,xie2017parallelizing}. } 
PI and PSL partition the data layer into disjoint key ranges, and assign each key range to a NUMA node along with the corresponding index layers. 
Each node is an array that allows cache locality and enables use of Single Instruction, Multiple Data (SIMD) instructions
to speed up comparisons during  search. Queries are  batch-processed. Incoming queries are grouped into batches  sorted by key. After a batch is formed, each query in the batch is routed to the NUMA node that operates over the query's key. PI and PSL support lock-free concurrency. 
A background thread reconstructs the index layers when the number of deletes and updates reaches a threshold. 
Search  is routed to the corresponding NUMA node before the in-NUMA search.
Insert  is routed to the corresponding NUMA node and is inserted into its skiplist.
When the number of inserts exceeds some threshold, the background thread propagates changes to the index layers.
Delete is  routed to the corresponding NUMA node, and the data item is marked deleted. When the number of logically deleted items exceeds a threshold, the background thread deletes the marked items in the data layer, and propagates the changes to the index layers.
If the range of the search covers multiple NUMA nodes, the query is split into  disjoint sub-queries for the corresponding NUMA nodes and the results are combined later. 
When the number of updates and deletes exceeds a threshold, the index layers are  reconstructed.
If data access is skewed, PI uses self-adjusted threading by spawning threads proportional to the number of queries that the NUMA node has. When the number of threads uses all hardware threads, the NUMA node  offloads some operations to other NUMA nodes.
PSL is compared with three in-memory tree indexes; Masstree~\cite{mao2012mass}, FAST~\cite{kim2010fast} and ART~\cite{leis2013adaptive}. PSL shows the highest insert and point read performance, and good scalability similar to FAST~\cite{xie2017parallelizing}.

\noindent
{\bf The Braided Skiplist~\cite{kim2022listdb}. }
NUMA effect is more severe in PM than in DRAM. ListDB~\cite{kim2022listdb} is a NUMA-aware braided skiplist
that  reduces  NUMA effects in PM.
The NUMA-oblivious skiplist may have data items scattered across NUMA nodes. Thus, search involves both local and remote NUMA access. The braided skiplist is a skiplist invariant where each upper layer is a sub-list of the bottom layer, and the upper layer does not have to be a sub-list of its next layer. The upper layer pointers can only point to data within the same NUMA node; while the bottom layer is connected via local and remote pointers sp that any data item can be found starting from any skiplist.

\subsection{The Graphics Processing Unit (GPU) Environment}\label{subsec:gpu}
GPUs are  crucial in databases as they offer massive parallelism and high memory-bandwidth.

\subsubsection{GPU-based Skiplist Variants}\hfill

\noindent

\textbf{GFSL~\cite{moscovici2017gpu}.}
GFSL is a GPU-friendly lock-based skiplist that improves data locality that is crucial for GPU computation. Multiple keys are grouped into cache-aligned chunks. Each chunk or skiplist node is equipped with a lock and a next pointer. Threads are divided into teams, same or smaller than a warp-size. A warp of threads executes the same instruction. The number of entries per chunk is the same as the number of threads in a team. During search, a team of threads reads the entire chunk with each reading one entry. Next, the team of threads uses intra-warp operations to decide to continue the same level or descend to the next level.
GFSL is compared with the skiplist algorithm ported to GPU in~\cite{prabhakar2012gpu} under different combinations of inserts, deletes and point reads, as well as different sizes of the key range. GFSL shows better performance with larger key ranges, and has superior performance for range sizes over 30k~\cite{moscovici2017gpu}.

\begin{figure*}[h]
    \centering
    \hspace*{\fill}%
    \begin{subfigure}{0.2\textwidth}
    \centering
        \includegraphics[width=0.8\linewidth]{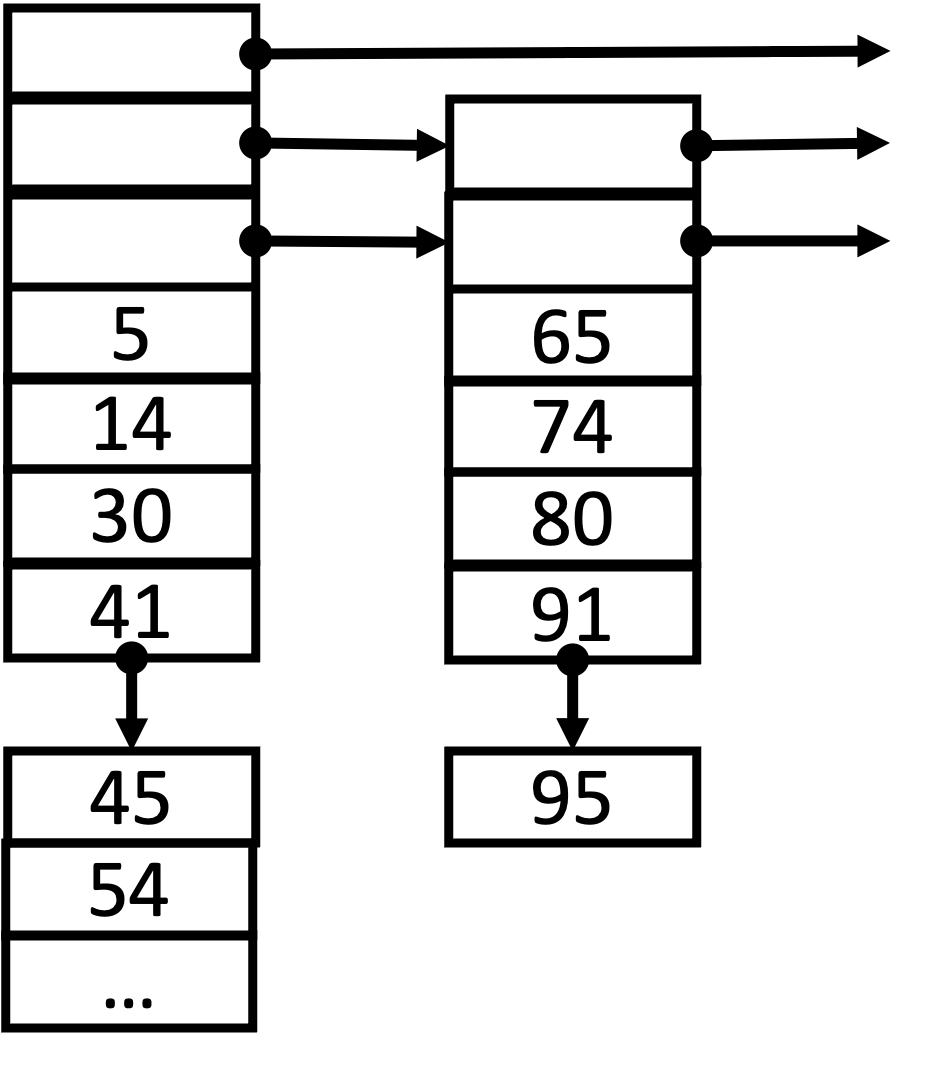}
        \caption{Structure of CMSL}\label{fig:cmsl-1}
    \end{subfigure}
    \hfill
    \begin{subfigure}{0.2\textwidth}
    \centering
        \includegraphics[width=0.8\linewidth]{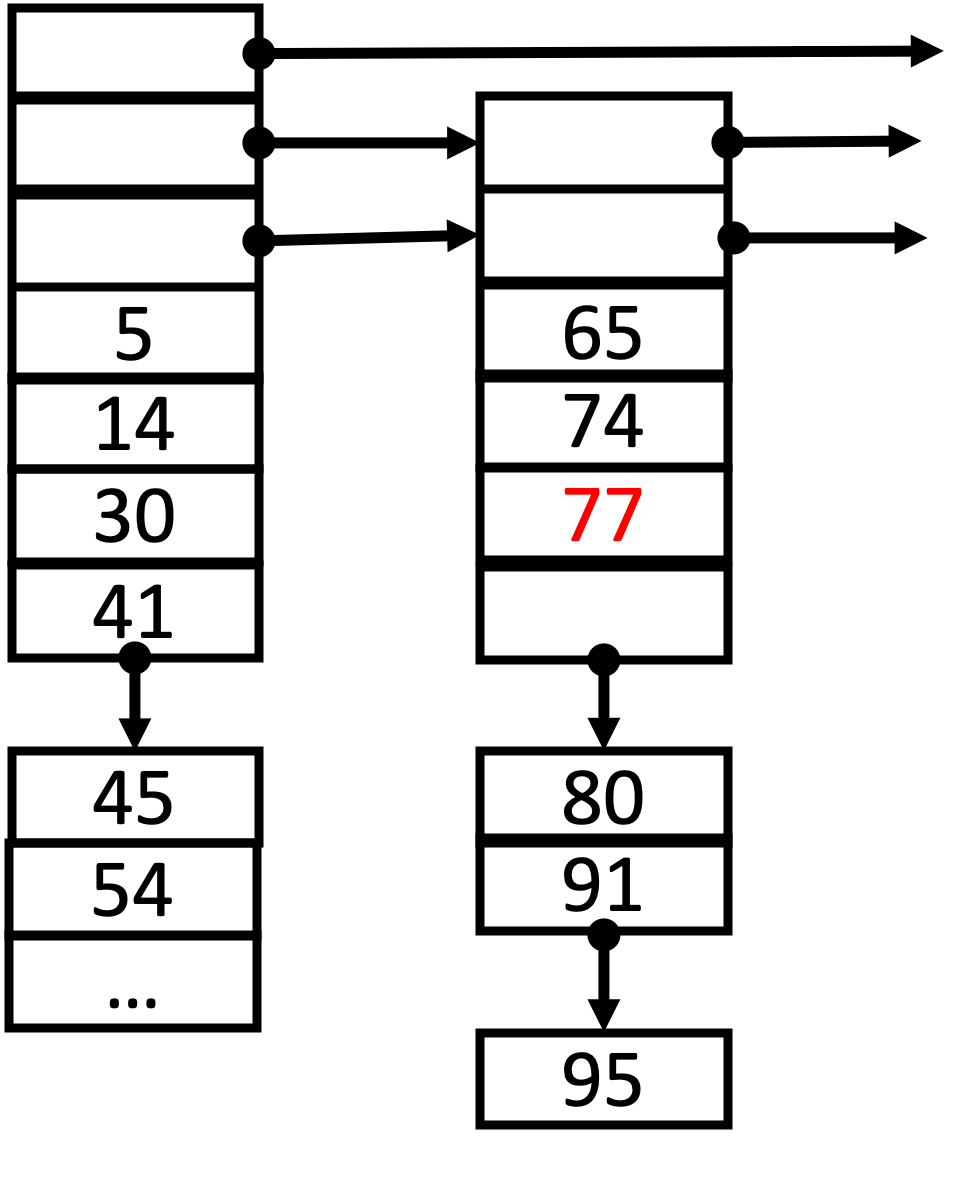}
        \caption{Insert 77 at level 0}\label{fig:cmsl-2}
    \end{subfigure}
    \hfill
    \begin{subfigure}{0.3\textwidth}
    \centering
        \includegraphics[width=0.78\linewidth]{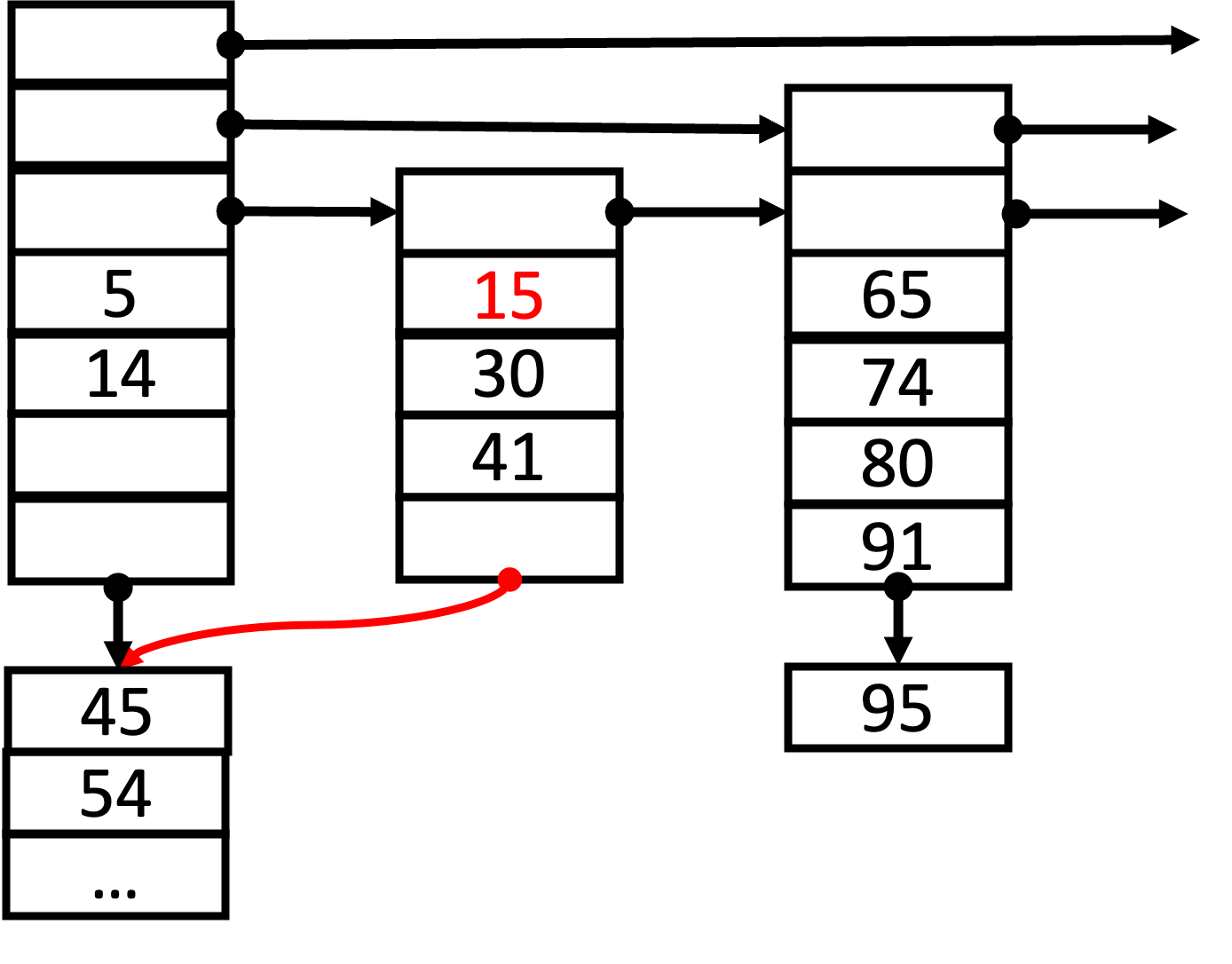}
        \caption{Insert 15 at Level 1}\label{fig:cmsl-3}
    \end{subfigure}
    \hspace*{\fill}%
    \caption{The structure of CMSL and its insert process.}\label{fig:cmsl}
\end{figure*}

\noindent

\textbf{CMSL~\cite{gpulockfree}.}
\underline{C} for \underline{M}edia \underline{S}kip\underline{L}ist (CMSL) is designed for Intel's on-die GPU (iGPU) using the C for Media framework. Sorted keys are kept in chunked lists that allow SIMD operations  (Figure~\ref{fig:cmsl-1}). The chunk of levels can be an 8d, 16d, 24d, or 32d word vector. The chunk of keys is a 16d word 
vector. 
This helps read at the full bandwidth during search. Search for the chunk list is followed by search within the chunk. For insert, there are two situations. If the random level is 0 for the inserted key, e.g., 77, the key is inserted into the chunk, and  the chunk may get split if necessary (Figure~\ref{fig:cmsl-2}). If the random level is greater than 0, e.g., 1, is assigned to a key, e.g., 15, a new list is created that includes a one-level chunk and new chunks of keys. Data from the old list, e.g., 30 and 41, are copied after 15, and the pointer to the next chunk, e.g., 45 and  54, is atomically set. CMSL is compared with CPU-based lock-free and lock-based skiplists. CMSL outperforms under all combinations of ratios of searches-inserts-deletes. When the workload is read-heavy, the speedup is about 3.1$\times$; when all the operations are updates, the speedup is 1.2-1.5$\times$.

\subsubsection{Performance Evaluation}

\cite{prabhakar2012gpu} compares  the lock-free linked-list, the lock-free hash table, and the lock-free skiplist  on GPU~\cite{herlihy2020art}. Two workloads are used (1)~20\% inserts, 20\% deletes, 60\% lookups; and (2)~40\% inserts, 40\% deletes, 20\% lookups. The GPU-based lock-free skiplist has significant increase in speedup for  small and medium key ranges. For large key ranges, the speedup drops significantly due to the overhead of the atomic operations. For the lesser lookup workload~(2) above, the speedup is more significant~\cite{prabhakar2012gpu}. The lock-free skiplist is also compared with the lock-free linked list. The performance gain of the skiplist over the linked list is larger on  GPU~\cite{prabhakar2012gpu}.

\subsection{Near-memory Processing}\label{subsubsec:nmp}

Near-memory processing (NMP, for short) 
alleviates the 
memory bottleneck by bringing computation closer to or directly inside the memory.

\noindent

{\bf The NMP-skiplist~\cite{liu2017nmp,choe2019nmp}.}
Main-memory consists of DRAM that is accessible from CPUs directly as well as the PIM memory that is divided into partitions, termed PIM vaults~\cite{liu2017nmp}. The flat-combining technique~\cite{hendler2010flat} (ref Section~\ref{subsec:concurrent-skiplists}) can be applied such that one CPU thread becomes the combiner to execute the requests in the publication list. However, as the CPU sends requests to the local NMP core to execute, the combiners require locks on the skiplist. When two requests access nodes that do not have large overlapping sub-paths, the search within the PIM vault is less effective because of the global locking~\cite{liu2017nmp}. 
Thus, a skiplist is partitioned into disjoint ranges. Each partition starts at a sentinel node that is of the maximum height. A partition covers the range from the key of the current sentinel node to the one of the next partition. CPUs store the keys of the sentinel nodes in regular DRAM. Search starts by comparing the key to the sentinel nodes, and continues in the specific partition to retrieve the value.
\cite{liu2017nmp} compares the proposed skiplist with the flat-combining version. Both techniques show increased throughput with more partitions, and the NMP-managed skiplist is better than the flat-combining one with the same number of partitions~\cite{liu2017nmp}.

\noindent
{\bf HybriDS~\cite{choe2022hybrids}.}
Choe et al.~\cite{choe2022hybrids} present a hybrid skiplist for near-memory processing. The hybrid structure is managed by two sets of threads: Host CPU and near-memory compute unit and the total accessible main memory includes host-accessible memory and NMP-capabable memory.
The data structure is split into two parts managed by these two sets of threads. With the hierarchical nature of the skiplist, the upper levels are placed in the host-managed portion to have an approximate size of the last-level cache, and the remaining levels are kept in NMP cores. 

Flat-combining~\cite{hendler2010flat}, described in Section~\ref{subsubsec:lock-based}, is adopted to support concurrency in the NMP portion.
The NMP core is single-threaded, and the combiner thread iterates over the list to check for the unserved requests. Inserts are applied to the NMP-managed portion first while deletes are applied to the host-managed portion first.
The hybrid skiplist is compared against an NMP-based skiplist~\cite{choe2019nmp} and a lock-free non-NMP~skiplist\cite{fraser2004practical}. The hybrid skiplist outperforms its opponents under the YCSB benchmark with different numbers of threads, and has relative advantage with more concurrent operations~\cite{choe2022hybrids}.

\section{Skiplist-Based Key-Value Stores}\label{sec:kvstore}
We highlight several  deployed key-value store systems that  utilize skiplists to map  keys to their corresponding values.
The advantages of the skiplist are that they keep the keys sorted, and hence facilitate fast search, as well as allow for fast inserts and deletes as well as range searches. Moreover, they allow for high concurrency. Specifics of how certain key-value store systems make use of the skiplist data structure is given below.
\textit{SingleStore}, previously known as MemSQL~\cite{memsql}, uses the skiplist as an index for its row-stores. Nitro~\cite{lakshman2016nitro} uses a lock-free skiplist~\cite{sundell2005fast} as its core index. We have discussed its MVCC layer in Section \ref{subsubsec:mvcc}. ListDB~\cite{kim2022listdb} is an LSM-Tree based key-value store for Persistent Memory (PM). RS-store~\cite{huang2021rs} is a skiplist-based key-value store using RDMA.

\noindent
{\bf RS-store~\cite{huang2021rs}.}
Remote Direct Memory Access (RDMA, for short) is a high-speed network communication mechanism with low latency that allows direct access to memory in remote nodes~\cite{kalia2016design}. RS-store~\cite{huang2021rs} is a skiplist-based key-value store using RDMA. RS-store's skiplist, termed the R-skiplist, follows a block-based approach, where it groups multiple skiplist entries into one block. This saves communication cost as one communication can fetch a block instead of one node of the skiplist. For increased scalability, concurrency control is required on the server-side as well as between the client and the server in the R-skiplist. On the server side, an exclusive access strategy is adopted to omit latches. The R-skiplist and tasks are partitioned into multiple partitions, where each partition is accessed exclusively by one thread. Write operations that are only performed on the server side are serialized within one partition. Since partitions are generated based on the top-level nodes, more than one thread can visit the same top-level block causing conflicts. 
Conflicts are 
resolved by a latch-free reference counting technique (Refer to~\cite{huang2021rs} for further detail). For controlling concurrency between the server and the client, remote read atomicity is guaranteed by a verification-based technique. A flag on the block is set before every write operation, and a remote read verifies the flag before reading the fetched block. The R-skiplist supports range search without copying and sending data from the server to the client. After the client issues a range search request, the server returns once with the starting address of the block, and the client can issue remote RDMA reads to retrieve the  next blocks.
RS-store is compared with RocksDB~\cite{rocksdb} that is modified to support RDMA~\cite{huang2021rs}. RS-store outperforms RocksDB in both local and remote access. RS-store has a higher server CPU utilization ratio~\cite{huang2021rs}.

\noindent
{\bf ListDB~\cite{kim2022listdb}.}
ListDB~\cite{kim2022listdb} proposes a skiplist that is NUMA-aware on persistent non-volatile memory (PM, for short)~\cite{izraelevitz2019basic}, and uses a skiplist to build a key-value store. ListDB is a three-level LSMT with one volatile level in DRAM and L0 and L1 levels on PM. During data flushing from DRAM to L0, ListDB restructures the log entries in PM as a skiplist. Thus, write amplification is reduced. Once the pointers between skiplist nodes are persisted via \verb|clflush| (details in Section~\ref{subsubsec:PM}), this skiplist is now in L0. Later, this L0 skiplist can be merge-sorted with the existing skiplist in L1 in-place.
ListDB is evaluated against PM tree-based indexes, namely, the BzTree~\cite{arulraj2018bztree}, the FPtree~\cite{oukid2016fp}, FAST-FAIR~\cite{hwang2018fair} and the PACTree~\cite{kim2021pactree}, and LSM-based key-value stores, namely,
NoveLSM~\cite{kannan2018novelsm}, SLM-DB~\cite{kaiyrakhmet2019slmdb}, Pmem-RocksDB~\cite{pmemrocksdb}. Under write-intensive workloads, ListDB outperforms the tree-structured indexes~\cite{kim2022listdb}. Under read-only workloads, ListDB shows increased throughput with larger lookup cache size, and can outperform or show comparable performance to those for the tree-structured indexes. ListDB outperforms the LSM-based key-value stores in the write-only workload, and the read throughput is slightly worse unless the lookup cache is enabled~\cite{kim2022listdb}.

\section{Indexing Complex Data}\label{sec:complex-data}
The skiplists described so far focus on indexing one-dimensional data, e.g., numbers and strings. The skiplist has been extended to index multi-dimensional data and other types of data.

\subsection{The Multi-dimensional Skiplist}\label{sec:multi-dim}
Similar to its tree-based counterparts, skiplists  can also be used to index multi-dimensional data. In this section, we discuss  skiplists that can be used to index multi-dimensional data. We assume that there are $n$ total points, and each point is in the $k$-dimensional space.

\noindent
{\bf The $k$-d Skiplist~\cite{nickerson1998skip}.}
Skiplists  have been extended to support multi-dimensional data and three versions of the skiplists have been introduced in~\cite{nickerson1998skip}.
\textit{Version 1}:
This version uses projection to represent all data points. Multiple $k$ deterministic skiplists are created, one for each dimension. To insert a point, the corresponding value of the $i^{th}$ dimension is inserted in the $i^{th}$ skiplist. Each node points to the memory location of the point $p$. 
\textit{Version 2}:
In the second version, there are $k$ skiplists, one for each dimension.
In addition, each node has $k$ additional dimensional pointers that $(k - 1)$ of them point to the same node's corresponding location in the other $(k - 1)$ skiplists, and one point to itself.
\textit{Version 3}: 
In this version, a point is only stored in one node instead of storing the same point in $k$ different nodes (one in each skiplist of each dimension). This considerably reduces memory usage. One disadvantage of this approach is that if the points are not uniformly distributed in all the dimensions, then the multi-dimensional skiplist degenerates into a single skiplist.

\noindent
{\bf The $k$-d Range Deterministic Skiplist (The $k$-d Range DSL)~\cite{lamoureux2005deterministic}.}
This skiplist is designed for $k$-d range search. Its basic structure is similar to the 2-3-4 tree in Figure~\ref{fig:skiplist-tree}. In Dimension 1 of the 2-d Range DSL, a node contains the following entries: Minimum value of the subtree below, Maximum value of the subtree below, the right pointer, the low pointer, the data (only bottom level/leaf node contains data), the \textit{nextDim} pointer~\cite{lamoureux2005deterministic}. As there are multiple dimensions, the \textit{nextDim} pointer points to Dimension 2 of the subtree below this node~\cite{lamoureux2005deterministic}.

\noindent
{\bf The Skip QuadTree~\cite{eppstein2005skip}.}
QuadTrees index multi-dimensional data by partitioning the space into quadrants. 
Quadtrees are unbalanced and can have depth $O(n)$, and can have poor update and search times in the worst case. Eppstein et al.~\cite{eppstein2005skip} combine a skiplist and a compressed Quadtree to achieve a balanced structure that can be used to answer point queries, approximate range count, and approximate nearest-neighbor search queries. 

\subsection {The Interval Skiplist}\label{sec:interval}
The interval skiplist~\cite{hanson1991interval} is another skiplist variant. Given $n$ intervals $(a_1,b_1)$, $(a_2,b_2)$,  $\cdots$, $(a_n,b_n)$ and a number $x$, the interval skiplist returns all the intervals that overlap $x$ in $O(\log n + L)$ time, where $L$ is the number of matching intervals. 
Given an interval $I = (a, b)$, the interval skiplist places Nodes $a$ and $b$ into the skiplist. The interval skiplist places marker `I' to represent the interval on forward edges using containment and maximality rules~\cite{hanson1991interval}. A containment rule ensures that a marker `I' will be placed on a forward edge going from X to Y, if I contains interval (X,Y). A maximality rule ensures that there is no forward edge corresponding to an interval (X', Y') that lies within I and contains (X,Y). Moreover, a node X will have markers for all intervals I, if X $\in$ I.

\begin{figure}[h]
\centering
        \includegraphics[width=0.5\linewidth]{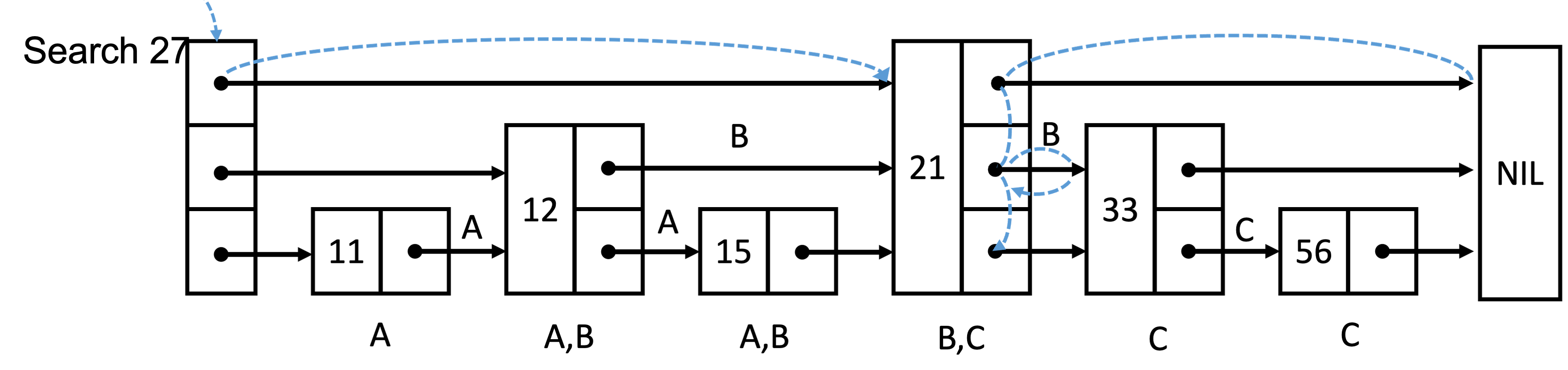}
\caption{The interval skiplist~\cite{hanson1991interval} corresponding to intervals $A = [11, 15], B = [12,33), C = [21, 56]$ and search path (dashed lines) for finding all intervals that overlap with 27.}\label{fig:interval-skiplist}
\end{figure}

Figure~\ref{fig:interval-skiplist} shows the interval skiplist for the three intervals $A = [11, 15]$, $B = [12,33)$, $C = [21, 56]$. To find all intervals that contain the number 27, the interval skiplist is searched for 27, and the union of all the markers in edges along the search path is computed as well as the marker of the last node that the search is stopped at. The search path of 27 is shown in dashed lines in Figure~\ref{fig:interval-skiplist}. Let $S$ be the set containing all the intervals that overlap the Number 27. Initially,  $S$ is empty, after traversing the forward edge from Node 21 to Node 33, $B$ is added to $S$, and finally the search terminates, all markers of Node 21 are added to the set $S$. The resulting set is $\{B, C\}$, so Number 27 overlaps Intervals $B = [12, 33)$ and $C = [21, 56]$.
The interval skiplist~\cite{hanson1991interval} has been realized in the Ariel Active Database system~\cite{hanson1996design}. Later,  the interval skiplist has been shown to perform similarly to tree-based alternatives~\cite{hanson1996selection}.

\section{Supportive Role of the Skiplist}\label{sec:support}
Many LSM-Tree Key-Value stores use a skiplist to store the in-memory data for fast concurrent insert while maintaining data in sorted order, e.g., LevelDB~\cite{ghemawat2011leveldb}, RocksDB~\cite{rocksdb}, X-engine~\cite{huang2019x}, Redis~\cite{redis}, and HBase~\cite{hbase}. Also, skiplists have  been used in combination with other data structures to improve the scalability of the in-memory component of key-value stores, e.g., S3~\cite{zhang2019s3}, FloDB~\cite{balmau2017flodb}, and TeksDB~\cite{han2019teksdb}. We present how the skiplist is modified to better integrate with LSM-Trees (Section~\ref{subsec:integrating}). Then,  we  show how skiplists can be used as priority queues (Section~\ref{subsec:priority-queue}).

\subsection{Integrating Skiplist with LSM-Trees}\label{subsec:integrating}
Log Structure Merge Trees are used in NoSQL applications and databases, e.g., in the database systems LevelDB~\cite{ghemawat2011leveldb}, RocksDB~\cite{dong2021evolution}, and FloDB~\cite{balmau2017flodb}. LSM has multiple levels, each with exponentially increasing memory capacity. All the arriving data, e.g., key-value pairs, are first sent to the LSM's in-memory component. When the in-memory component gets filled, the key-value pairs are compacted into a block format, and then is flushed into permanent storage afterwards. The choice of data structure for the memory component of many LSM realizations is the skiplist, as it is a lightweight data structure that maintains the data in sorted order. Operations, e.g.,  insert, delete, find, and range search can be performed in $O(\log n)$ time. Once the main-memory component that stores the skiplist becomes full, it is persisted in stable storage. LevelDB and RocksDB use Probabilistic Skiplists. In RocksDB, skiplists support concurrent inserts and lookups.

\noindent
{\bf FloDB~\cite{balmau2017flodb}.}
An LSM-tree data store has an in-memory and disk components. Operations are  executed in the in-memory component, and then in the subsequent levels on disk. Popular LSM-tree implementations, e.g., LevelDB~\cite{ghemawat2011leveldb} and RocksDB~\cite{rocksdb} use a skiplist as their in-memory component. FloDB~\cite{balmau2017flodb} has a custom two-level in-memory structure that improves performance as the memory size increases. 
The in-memory component can either be sorted, e.g., as in a skiplist, or unsorted, e.g., as in a hashtable. 
Experiments~\cite{balmau2017flodb} show that the hashtable is good at dealing with write-heavy workloads, whereas skiplists are good at dealing with read- and scan-heavy workloads. 
FloDB's uses a concurrent hash-table, termed the membuffer, and a skiplist, termed the memtable. This two-level index in main-memory gets the best of both structures.
The membuffer (hash-table) is smaller in size compared to the memtable (skiplist). The hash-table takes the $l$-most significant bits as a hash function. Whenever the membuffer is full, it is compacted into the memtable. And when the memtable is full, it is flushed into disk. 
Point lookup checks the membuffer. If not found, the memtable and the disk may be searched accordingly.
A new data item is  inserted into the membuffer. If the membuffer is full, then inserts are directed into the memtable. 
Since range search is faster in the skiplist, data is flushed from the membuffer into the skiplist to perform  range scans in the skiplist.
FloDB allows bulk inserts to insert batches of key-value pairs in ascending order into the skiplist. FloDB~\cite{balmau2017flodb} stores the predecessor node's path to speed up insert time for a node. It benefits from neighborhood proximity to improve speed. Also,  it supports concurrency. 
In~\cite{balmau2017flodb}, FloDB is compared against LevelDB~\cite{ghemawat2011leveldb}, RocksDB~\cite{rocksdb} and HyperLevelDB~\cite{hyperleveldb}. For write-only workloads, FloDB outperforms   HyperLevelDB (the second best) by 1.9$\times$ to 3.5$\times$. For read-only workloads, RocksDB outperforms FloDB when the number of threads is over 16. In three mixed workload situations: 50\% reads, 25\% inserts, 25\% deletes; one writer thread and the remaining are read threads; 95\% updates, 5\% range searches, FloDB outperforms all other systems for up to 16 threads.

\noindent
{\bf S3~\cite{zhang2019s3}.}
Similar to FloDB, S3 proposes a novel in-memory data structure that improves the efficiency of the overall database. S3 is seamlessly integrated with the disk part of LevelDB and RocksDB, and is comparable in speed to the Adaptive Radix Tree (ART)~\cite{herlihy2020art}. S3 uses the Fast Architecture Sensitive Tree (FAST)~\cite{kim2010fast}, an index tree that fits in the cache on top of a semi-ordered skiplist. FAST uses neural networks to predict the guard entries, and achieves comparable throughput to that of ART. 
Search starts from the root of FAST,
and finally reaches a guard entry in a skiplist. Then, it proceeds to find the guard range in which the searched key 
is located using  regular search in of the  skiplist. 
After locating for the proper place to insert an item, it is placed within its guard range. 
In~\cite{zhang2019s3}, S3 is integrated with RocksDB~\cite{rocksdb} to be compared with the original RocksDB in the \texttt{db\_bench} benchmark provided by RocksDB. The integrated RocksDB+S3 shows higher insert and lookup throughputs for up to 16 threads. S3 is also compared with other in-memory indexes, including Cicada~\cite{lim2017cicada}, the Masstree~\cite{mao2012mass}, and the Bwtree~\cite{levandoski2013bw}. S3 has the highest throughput under both insert and lookup workloads with uniform and skewed data distributions~\cite{zhang2019s3}.

\noindent
{\bf TeksDB~\cite{han2019teksdb}.}
Han et al.~\cite{han2019teksdb} observe that no single data structure can achieve both fast point access and fast range search at the same time. Also, simple combination of multiple data structures that are each optimal for one operation cannot work~\cite{han2019teksdb}. Thus, TeksDB proposes a hybrid data structure by ``judiciously weaving" two complementary structures: a CuckooHash~\cite{pagh2004cuckoo} and a skiplist. CuckooHash is responsible for point lookups and updates while the skiplist serves range searching. To save space and avoid duplicating data among the two data structures, the key-value pairs are kept in a pool, and the two data structures only keep pointers to this pool.
In~\cite{han2019teksdb}, TeksDB is compared against key-value stores, namely, Redis~\cite{redis}, WiredTiger~\cite{wiredtiger}, LMDB~\cite{lmdb} and RocksDB~\cite{rocksdb} using fit-in-memory dataset with the YCSB benchmark. TeksDB outperforms the other systems in Workloads A, B, C, D and F and is slower than Redis in loading~\cite{han2019teksdb}.

\subsection{The Skiplist as a Priority Queue}\label{subsec:priority-queue}
A priority queue offers two operations: the \texttt{Insert}, and \texttt{DeleteMin} operations. The skiplist is already ordered, and hence can be modified easily to behave as a priority queue. 
In this section, we overview several skiplist implementations of the priority queue.

\noindent
\textbf{The Lock-Based Priority Queue~\cite{shavit2000skiplist}.}
In the lock-based priority query~\cite{shavit2000skiplist}, the skiplist is extended using lock-based concurrent skiplists~\cite{pugh1998concurrent}. Each node has a flag that is set to false while inserting and is set to true when logically deleted.  Inserts execute in  the same way as in the concurrent skiplist. \texttt{DeleteMin} traverses the base linked list and finding the first node that is not marked as deleted. Then, the process  locks and marks the node as deleted, and then is  physically deleted. This allows extending the skiplist into a priority queue. Because the skiplist is a distributed structure and does not require rebalancing, it is a good candidate to act as a concurrent priority queue with multiple processors. 
\cite{hendler2010flat} compares the lock-based priority queue~\cite{shavit2000skiplist} with a flat-combining one (Section~\ref{subsubsec:lock-based}). The workload is 50\% inserts and 50\% \texttt{DeleteMin}. 
The flat-combining priority queue outperforms the lock-based priority queue by up to 3$\times$~\cite{hendler2010flat}.

\begin{figure*}[h]
\centering
\hspace*{\fill}%
    \begin{subfigure}{0.5\textwidth}
    \centering
        \includegraphics[width=\linewidth]{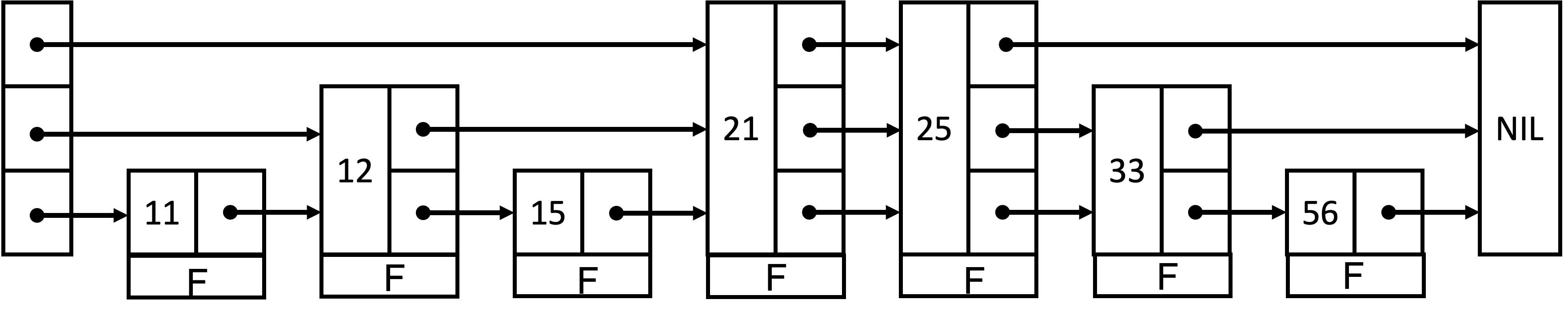}
        \caption{One example of skiplist}\label{fig:pq_skiplist}
    \end{subfigure}
    \hfill
    \begin{subfigure}{0.22\textwidth}
    \centering
        \includegraphics[width=0.75\linewidth]{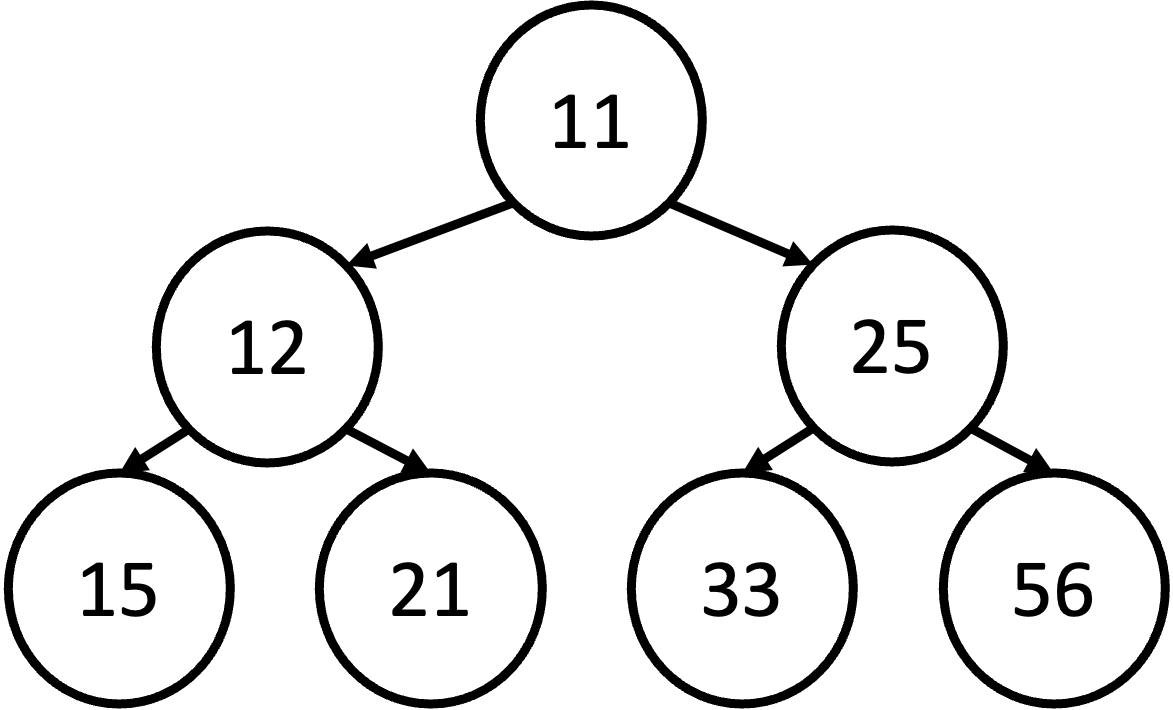}
        \caption{Priority queue }\label{fig:pq_pq}
    \end{subfigure}
    \hspace*{\fill}%
\caption{A skiplist and its equivalent priority queue}\label{fig:Skiplist_as_PQ}
\end{figure*}


\noindent

{\bf The Lock-free Priority Queue~\cite{sundell2005fast}.}
The lock-free skiplist-based priority queue uses  two-step delete (Section~\ref{subsubsec:lock-free}). In \texttt{DeleteMin}, a node is logically deleted followed by a physical delete. To ensure linearizability, only one single logically deleted node can exist at the bottom level. The subsequent thread will help in the physical removal of the node if it finds one in the bottom level~\cite{sundell2005fast}.
A proof is  given in~\cite{sundell2005fast} that the implementation is correct and is linearizable.

\noindent

{\bf The Linearizable Priority Queue~\cite{linden2013skiplist}.}
The bottleneck of a lock-free priority queue~\cite{sundell2005fast} in a concurrent environment mainly comes from the physical delete of the node during the \texttt{DeleteMin} operation. In~\cite{linden2013skiplist}, \texttt{DeleteMin} does not  physically delete a node. New nodes can still be added before the first non-deleted nodes and the logically deleted nodes are physically removed in batches~\cite{linden2013skiplist}.
This priority queue is compared against Fraser's skiplist implementation~\cite{fraser2004practical}, the 
lock-free one~\cite{sundell2005fast} 
and a lock-free adaptation~\cite{herlihy2020art} of~\cite{shavit2000skiplist}. The workload includes inserts and \texttt{DeleteMin} operations. The proposed priority queue is 30\% to 80\% faster. The performance scales until 8 threads where a steep drop is observed that is due to the inter-socket communication with more threads~\cite{linden2013skiplist}.

\noindent
{\bf The Spraylist~\cite{alistarh2015spraylist}. } 
The spraylist~\cite{alistarh2015spraylist} is another priority queue implementation using a lock-free skiplist and a custom ``spray'' function to achieve a concurrent relaxed priority queue that scales well in high-contention scenarios. The spraylist's \verb|DeleteMin|   returns an element among the first $O(q\log^3q)$ elements in the list with high probability, where $q$ is the number of threads. This relaxed concurrent priority queue reduces the contention among threads to compete for the minimal node.  \verb|DeleteMin|  uses Operation \verb|spray| that
starts at the front of the skiplist (the header) and at some initial height $h = \log q + K$ where $K$ is a constant,
performs a random walk that goes right horizontally and down vertically to the bottom level, to find the returned node. Dummy nodes are padded in the front of the spraylist such that the very first few nodes can also be returned by the \verb|spray| operation. Detailed description of the spraylist and discussion of its parameters can be found in~\cite{alistarh2015spraylist}.

The spraylist is compared against the lock-based one~\cite{shavit2000skiplist} (L1), the linearizable one~\cite{linden2013skiplist} (L2), and a priority queue that uses~\cite{fraser2004practical} (L3). L1 and L2 do not scale with more threads while L3 performs the best and is better than SprayList by a constant factor~\cite{alistarh2015spraylist}.

\noindent
{\bf Tle Flat-Combining Skiplist-Based Priority Queue~\cite{hendler2010flat}.}
A skiplist-based priority queue is introduced in~\cite{hendler2010flat}. To implement \texttt{removeSmallestK}, the bottom level is traversed, and the first $k$ items are returned and are removed while adjusting the skiplist structure accordingly. To implement \texttt{combinedAdd}, a list of sorted items <$i_1$, $i_2$, ..., $i_k$> are added in one pass. To insert $i_1$, the search for $i_1$ starts from the highest level. Once the search arrives at a location where the predecessor and successor of $i_1$ and $i_2$ differ, that location is recorded and the search for $i_1$ proceeds. Once $i_1$ is 
inserted to its destination location,
the search for $i_2$ is resumed starting at the recorded location.

\section{Efficient Range Search with the Skiplist}\label{sec:range-query}

The cache-sensitive skiplist~\cite{sprenger2017cache} has been presented in Section~\ref{subsec:cache}. It adopts an unrolled node structure to improve  cache locality and range search performance.
We 
overview
another skiplist below that also supports efficient range search.

\noindent
{\bf The Leap-List~\cite{avni2013leaplist}.}
The leap-list is designed for linearizable range queries. Each node can hold up to $k$ immutable items. If $k$ is large, a bitwise trie is embedded within the node to facilitate 
lookup queries. Software-Transactional-Memory (STM) is used to implement short transactions that acquire locks. A range query starts by finding the predecessor of the node where the range query starts. Then, the transaction begins, and if the node is marked live, the transaction traverses the bottom level, and returns the result of the range query~\cite{avni2013leaplist}. In the experiments, the leap-list is compared with the skiplist~\cite{fraser2004practical}. 
For write-only workloads, the throughput of the skiplist is better than the leap-list by up to 10$\times$ for 80 threads. 
However, with more lookup and in range search-only workloads, the leap-list is 35$\times$ better than the opponent~\cite{avni2013leaplist}.

\section{The Extensibility of Skiplists}\label{sec:extensibility}
The principles and heuristics of skiplists can be applied across various domains. In this section, we present examples of extensions and adaptations of skiplists.

\noindent
{\bf PebblesDB~\cite{raju2017pebblesdb}.}
As the height of each inserted node is determined via randomness, the skiplist does not require rebalancing, which is different from the B$^+$tree. \textit{PebblesDB}~\cite{raju2017pebblesdb} combines the skiplist with Log-Structured Merge Trees (LSM-tree) in the context of the fragmented LSMT (FLSM). Ordinary LSMT flushes memory buffer to disk files, termed {\em sstable}s and compacts sstables in a hierarchical way~\cite{o1996log}. Since sstable is immutable, the same data can be written multiple times through its lifetime during compactions, thus data write amplification is high~\cite{raju2017pebblesdb}. FLSM adopts the probabilistic design of the skiplist in choosing which keys from the inserted keys in each level to split that level's key space. These chosen keys are termed {\em guards} as no files can span across guards.
Each guard has its associated sstables. 
Guards are chosen from the inserted keys in a similar way as to that of the skiplist. This design alleviates the issue of write amplification in the LSMT~\cite{raju2017pebblesdb}. 

In the write-amplication experiment~\cite{raju2017pebblesdb}, PebblesDB is the smallest among RocksDB~\cite{rocksdb}, HyperLevelDB~\cite{hyperleveldb}, and LevelDB~\cite{ghemawat2011leveldb}. In single-threaded experiments, PebblesDB outperforms its opponents in random writes, random reads and deletes, and is 3$\times$ less than HyperLevelDB in sequential writes, and has a low throughput in range searches~\cite{raju2017pebblesdb}. With four threads, PebblesDB performs the best under the write-only, read-only and concurrent read-write workloads~\cite{raju2017pebblesdb}.

\noindent
{\bf HNSW~\cite{malkov2018efficient}.}
Hierarchical Navigable Small World (HNSW) is used for K-Approximate Nearest Neighbour Search (K-ANNS)~\cite{malkov2018efficient}. This algorithm can be seen as an extension to the skiplist with proximity graphs replacing the linked list component in each level~\cite{malkov2018efficient}.

\begin{figure}[h]
    \centering
    \includegraphics[width=0.45\linewidth]{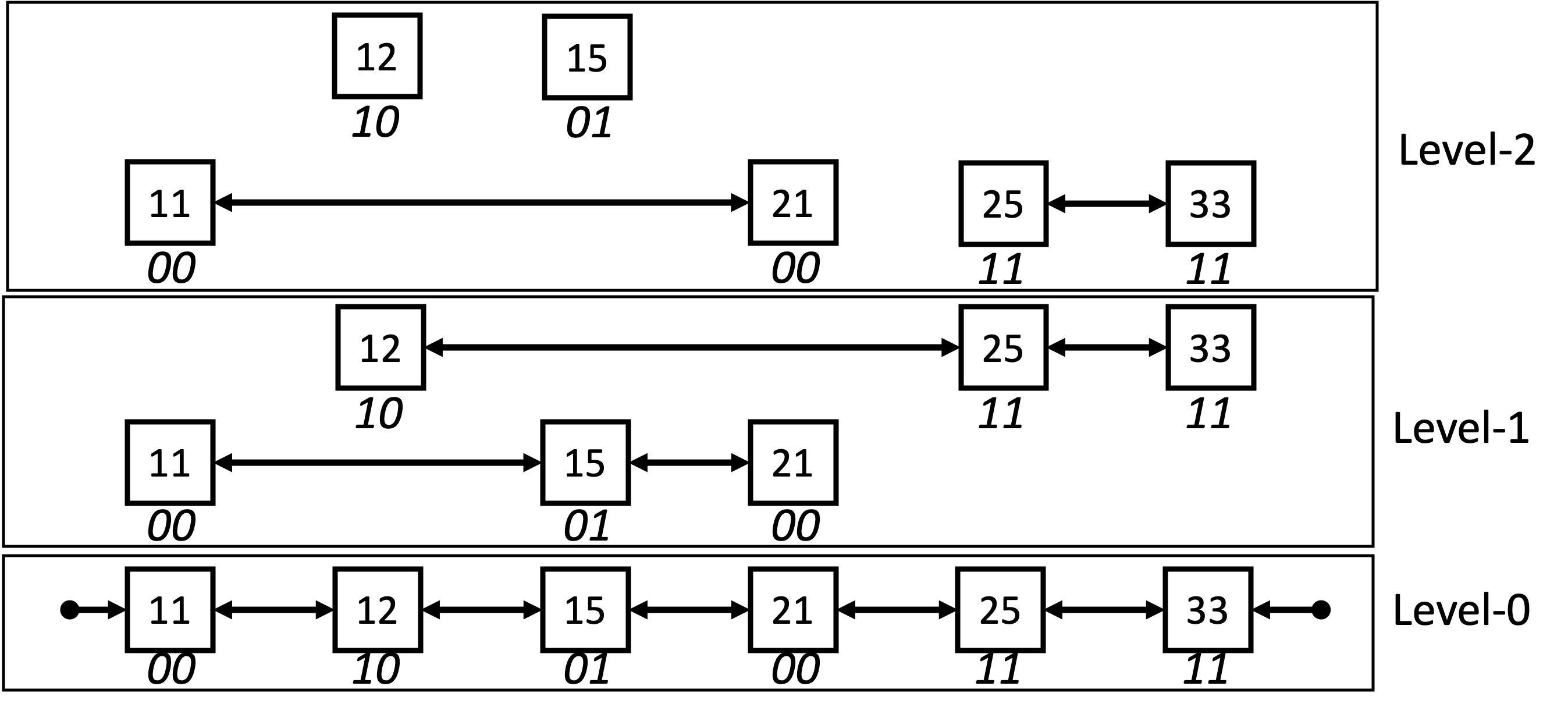}
    \caption{A skip graph with $\lceil O(\log N)\rceil = 3$ levels.}
    \label{fig:skipgraph}
\end{figure}

\noindent
{\bf The Skip Graph~\cite{aspnes2007skip}.}
The skip graph is a distributed data structure based on the skiplist. The list in a skip graph is doubly-linked. With $N$ data items, the number of levels is $O(\log N)$. However, in a skip graph, there can be many lists at one level. The bottom level (Level-0) contains all data items in a doubly linked list. The upper level contains the same amount of data items but links fewer data items in one linked list. Each node participates in one of the lists until the node becomes a singleton node that is not linked with any other nodes after $O(\log N)$ level elevation on average. Each data item $x$ is given a random word of membership vector $m(x)$. For example,  in Figure~\ref{fig:skipgraph}, $m(15) = 01$. At Level-$i$, all items that share the same prefix of Length $i$ are in the same list, e.g., 11, 15, and 21 have Prefix 0 at Level-1 while 11 and 21 have Prefix 00 at Level-2. Search and insert are similar as in the skiplist. The skip graph is used in peer-to-peer networks. It supports range search while distributed hash tables cannot. The skip graph has logarithmic insert and delete performance, and can tolerate node failure without being disconnected. 

\section{More Uses of Skiplists}\label{sec:more-use}
Skiplists have been applied to other domains. We list some of them below.

\noindent
{\bf Network Overlay Algorithms.}
Due to the balanced nature and the ability to efficiently search and update the skiplist, it is widely used in network overlay algorithms. Tiara~\cite{clouser2008tiara} uses a sparse 0-1 distributed skiplist, and applies a self-stabilizing algorithm for the skiplist. Nor et al. propose the  Corona algorithm~\cite{nor2013corona} that constructs a self-stabilizing deterministic skiplist in message-passing systems. A distributed deterministic 1-2 skiplist has  been proposed~\cite{mandal2012deterministic}, and has been later extended~\cite{singh2015concurrent} to support concurrency.

\noindent
{\bf Other uses.}
Skiplists have  been used in asynchronous video streaming~\cite{wang2006peer} and VCR interactions~\cite{wang2008dynamic} due to their distributed and balanced nature.
Next generation sequencing data can  make use of the skiplist~\cite{lee2019ngs}. The huge size of the genomic data is converted to a 3-dimensional plot. Data is compressed and is stored in a skiplist for fast comparison and search~\cite{lee2019ngs}.
The skiplist is  applied in blockchain systems~\cite{singh2023blockchain}. To support historical queries on multi-dimensional data, a deterministic skiplist is used to index the data with an extra layer on top to help with retrieving historical data, where the skiplist is implemented on the chain node, i.e., the smart contract rather than the Hyperledger Fabric storage system~\cite{singh2023blockchain}.
The inverted index can make use of a skiplist. In~\cite{paolo2005inverted}, a compressed deterministic   skiplist, termed the perfect skiplist~\cite{paolo2005inverted}, is embedded into the inverted index such that unnecessary documents can be skipped quickly in an inverted list~\cite{paolo2005inverted}.
The skiplist has been applied in the domain of forecasting queries~\cite{duan2007forecasting}. In~\cite{ge2008forecasting}, the skiplist is used because of its multiple levels of granularity and its fast data access. Each level of the skiplist is associated with a set of models to predict including the bottom level~\cite{ge2008forecasting}.

\section{Concluding Remarks and Future Work}\label{section:conclude}
The skiplist has become ubiquitous, and is being used extensively in big data systems. This paper presents a survey of many variants of skiplists, and addresses the motivation behind each variant, its tradeoffs, and its potential applications. 
We have presented a taxonomy for skiplist variants and the related big data systems that make use of these variants. We have reviewed  existing systems that use skiplists, and how skiplists are used in each of these systems. 

Multiple optimization techniques have been applied to the skiplist to make it more versatile to be applied in various domains. The linked list nature of skiplist makes it more capabable of supporting concurrency. And skiplist partitioning is more straightforward compared to its tree counterparts. A deterministic skiplist is introduced to overcome the disadvantages of the original probabilistic skiplist. Skiplist can be adaptive to the skewness of the data. The nodes of the skiplist can be modified to hold either more data or a fixed number of pointers.

With the above optimizations, skiplists can be adopted widely as a data storage structure in various hardware platforms, or provides data indexing in different systems. Skiplist can also be integrated with other data structures and extended to be used in other domains.

There are still open challenges on skiplist, e.g., whether machine learning can be incorporated into either skiplist construction or searching is unknown, whether skiplist can be blended with more data structures to combine strength from both sides, how to adapt skiplist to the new emerging hardware, how to achieve a balance between deterministic and probabilistic behavior, how to use skiplists with other complex data types etc. We envision skiplist can be applied to more big data systems because of its good scalability, simple implementation and various variant types. We hope that this survey serve as a guide to researchers, practitioners, and end-users.

\section{Acknowledgements}
Walid G. Aref acknowledges the support of the National Science Foundation under Grant Number IIS-1910216.

\bibliographystyle{acm}
\bibliography{references}  






\end{document}